\documentclass[conference]{IEEEtran}
\IEEEoverridecommandlockouts

\usepackage[ruled,linesnumbered]{algorithm2e}

\def\BibTeX{{\rm B\kern-.05em{\sc i\kern-.025em b}\kern-.08em
    T\kern-.1667em\lower.7ex\hbox{E}\kern-.125emX}}
    
\usepackage{CJKutf8}  

\usepackage{hhline}
\usepackage{graphicx}
\usepackage{xcolor}
\usepackage{pgfplots}
\usepackage{tikz}
\usepackage{comment}

\usepackage{filecontents}
\usepackage{threeparttable}

\usepackage{pifont} 
\usepackage{amssymb}
\usepackage{bbding} 
\usepackage{pifont}

\usepackage{makecell}
\usepackage{colortbl} 

\usepackage{fancyhdr} 
\usepackage{hyperref}

\usepackage{multirow}
\usepackage{cite}
\usepackage{amsmath,amssymb,amsfonts}

\usepackage{textcomp}
\usepackage{xcolor}

\usepackage{subfigure}
\usepackage{pgfplots}
\usepackage{filecontents}

\usepackage{graphicx,subfigure}

\definecolor{bblue}{HTML}{4F81BD}
\definecolor{rred}{HTML}{C0504D}
\definecolor{ggreen}{HTML}{9BBB59}
\definecolor{ppurple}{HTML}{9F4C7C}

\usepackage{listings}

\lstdefinelanguage[ECMAScript2015]{JavaScript}[]{JavaScript}{
basicstyle=\tiny,
  morekeywords=[1]{await, async, case, catch, class, const, default, do,
    enum, export, extends, finally, from, implements, import, instanceof,
    let, static, super, switch, throw, try},
  morestring=[b]` 
}

\lstdefinelanguage{JavaScript}{
basicstyle=\tiny,
  morekeywords=[1]{break, continue, delete, else, for, function, if, in,
    new, return, this, typeof, var, void, while, with},
  morekeywords=[2]{false, null, true, boolean, number, undefined,
    Array, Boolean, Date, Math, Number, String, Object},
  morekeywords=[3]{eval, parseInt, parseFloat, escape, unescape},
  sensitive,
  morecomment=[s]{/*}{*/},
  morecomment=[l]//,
  morecomment=[s]{/**}{*/}, 
  morestring=[b]',
  morestring=[b]"
}[keywords, comments, strings]

\lstalias[]{ES6}[ECMAScript2015]{JavaScript}

\definecolor{mediumgray}{rgb}{0.3, 0.4, 0.4}
\definecolor{mediumblue}{rgb}{0.0, 0.0, 0.8}
\definecolor{forestgreen}{rgb}{0.13, 0.55, 0.13}
\definecolor{darkviolet}{rgb}{0.58, 0.0, 0.83}
\definecolor{royalblue}{rgb}{0.25, 0.41, 0.88}
\definecolor{crimson}{rgb}{0.86, 0.8, 0.24}

\lstdefinestyle{JSES6Base}{
basicstyle=\small,
  backgroundcolor=\color{white},
  basicstyle=\ttfamily,
  breakatwhitespace=false,
  breaklines=true,
  captionpos=b,
  columns=fullflexible,
  commentstyle=\color{mediumgray}\upshape,
  emph={},
  emphstyle=\color{crimson},
  extendedchars=true,  
  fontadjust=true,
  frame=single,
  identifierstyle=\color{black},
  keepspaces=true,
  keywordstyle=\color{mediumblue},
  keywordstyle={[2]\color{darkviolet}},
  keywordstyle={[3]\color{royalblue}},
  numbers=left,
  numbersep=5pt,
  numberstyle=\tiny\color{black},
  rulecolor=\color{black},
  showlines=true,
  showspaces=false,
  showstringspaces=false,
  showtabs=false,
  stringstyle=\color{forestgreen},
  tabsize=2,
  title=\lstname,
  upquote=true  
}

\lstdefinestyle{JavaScript}{
  language=JavaScript,
  style=JSES6Base
}
\lstdefinestyle{ES6}{
  language=ES6,
  style=JSES6Base
}

\newenvironment{packeditemize}{
	\begin{list}{$\bullet$}{
			\setlength{\labelwidth}{4pt}
			\setlength{\itemsep}{0pt}
			\setlength{\leftmargin}{\labelwidth}
			\addtolength{\leftmargin}{\labelsep}
			\setlength{\parindent}{0pt}
			\setlength{\listparindent}{\parindent}
			\setlength{\parsep}{0pt}
			\setlength{\topsep}{1pt}}}{\end{list}}

\usepackage{blindtext}
\usepackage{etoolbox,xstring,mfirstuc,textcase}
\usepackage{booktabs}
\usepackage{pgfplots}

\usepackage{wrapfig}

\usepackage{tabularx,booktabs,makecell,multirow, caption}
\usepackage{enumitem} 

\newcolumntype{L}{>{\arraybackslash}X}

\usepackage{xcolor}
\usepackage{tikz}
\usepackage{pgfplots}

\definecolor{findOptimalPartition}{HTML}{D7191C}
\definecolor{storeClusterComponent}{HTML}{FDAE61}
\definecolor{dbscan}{HTML}{ABDDA4}
\definecolor{constructCluster}{HTML}{2B83BA}

\usepackage[utf8]{inputenc}
\usepackage{listings}
\begin{document}
\title{Is Your A\underline{I} Truly Yours? Leveraging \underline{B}lockcha\underline{i}n for Copyright\underline{s}, Provenance, and Lineage}


\author{
        Qin~Wang\IEEEauthorrefmark{1},
        Guangsheng~Yu\IEEEauthorrefmark{2}, 
        Yilin~Sai\IEEEauthorrefmark{3}, 
        H.M.N. Dilum Bandara\IEEEauthorrefmark{1},
        Shiping Chen\IEEEauthorrefmark{1} \\
        
\IEEEauthorrefmark{1}\textit{CSIRO Data61} $|$
\IEEEauthorrefmark{2}\textit{University of Technology Sydney}  $|$
\IEEEauthorrefmark{3}\textit{UNSW Sydney, Australia}
}

\maketitle

\begin{abstract}

As Artificial Intelligence (AI) integrates into diverse areas, particularly in content generation, ensuring rightful ownership and ethical use becomes paramount, AI service providers are expected to prioritize responsibly sourcing training data and obtaining licenses from data owners. However, existing studies primarily center on safeguarding static copyrights, which simply treat metadata/datasets as non-fungible items with transferable/trading capabilities, neglecting the dynamic nature of training procedures that can shape an ongoing trajectory. 
In this paper, we present \textsc{IBis}, a blockchain-based framework tailored for AI model training workflows. Our design can dynamically manage copyright compliance and data provenance in decentralized AI model training processes, ensuring that intellectual property rights are respected throughout iterative model enhancements and licensing updates.
Technically, \textsc{IBis} integrates on-chain registries for datasets, licenses and models, alongside off-chain signing services to facilitate collaboration among multiple participants. Further, \textsc{IBis} provides APIs designed for seamless integration with existing contract management software, minimizing disruptions to established model training processes. We implement \textsc{IBis} using Daml on the Canton blockchain. Evaluation results showcase the feasibility and scalability of \textsc{IBis} across varying numbers of users, datasets, models, and licenses.

\end{abstract}

\smallskip
\begin{IEEEkeywords}
AI, Blockchain, License, Provenance, Trust
\end{IEEEkeywords}


\section{Introduction}
\label{sec-intro}

Artificial Intelligence (AI) technologies~\cite{bommasani2021opportunities,zhao2023survey}  have increasingly permeated various aspects of daily life, spanning from information retrieval~\cite{zhu2023large,bonifacio2022inpars} to content generation~\cite{li2022pretrained,chen2024benchmarking}. Concurrently, AI service providers have made strides in commercializing their services. Nevertheless, as AI models rely on extensive datasets aggregated from diverse sources for training~\cite{carlini2021extracting,hoffmann2022empirical}, apprehensions have emerged regarding the potential infringement of copyrights~\cite{chu2024protect,vyas2023provable,yu2023codeipprompt} during the data acquirement and model training process. To uphold responsible and ethical AI practices~\cite{lu2023developing,lu2023operationalizing}, comply with regulations, and reduce legal liabilities, AI service providers must actively collaborate with data owners, including content creators and media industry stakeholders. Establishing licensing agreements~\cite{power1970licensing,benjamin2019towards} and obtaining consent before utilizing data for AI model training is a key element of this collaboration~\cite{contractor2022behavioral}.
Hence, there is a growing need for new frameworks addressing data provenance, lineage, and copyright compliance in the AI industry, tailored to its distinct needs and workflows.

However, addressing the concerns of AI data provenance and copyright compliance can be a nontrivial task, particularly when the entire training process occurs locally or within a black-box cloud service~\cite{siddarth2021ai}, limiting transparency for users. 
To bridge this gap, we harness the properties of blockchain technology, which offers a tamper-proof and trustworthy environment~\cite{li2022smart} to establish authenticity, provenance, and lineage~\cite{nguyen2023blockchain,xu2019book}.
Owing to its inherent characteristics of immutability and transparency, blockchain has garnered widespread recognition as a suitable technology for achieving regulatory compliance~\cite{zhang2019BC,scott2018evaluate,allena2022BC}. For instance, data recorded on the blockchain is digitally signed and inherently tamper-proof, thereby constituting an authentic and persistent record that accurately reflects an event(s) at a specific point in time. This makes blockchain a fitting candidate to address concerns related to data provenance and copyright compliance within the AI industry~\cite{ural2023survey,hussain2021artificial,liu2020blockchain}.

We have identified a series of functional challenges for such a blockchain-based compliance framework~\cite{kiayias2022sok,liu2023systematic,xiong2025regkyc}:
(i) The framework needs to be designed to seamlessly integrate with the existing workflow of AI model training~\cite{jia2021proof,lan2021proof,wang2024aiarena}. (ii) The framework should support continuous model retraining and fine-tuning with new datasets~\cite{wang2024responsibility}, allowing for the generation of updated models while maintaining data provenance and lineage. (iii) The framework should support mechanisms for license expiration and renewal, accommodating diverse business models employed by data owners. (iv) The ownership of datasets and models~\cite{somy2019ownership,yu2024maximizing}, along with all training actions, should be accompanied by evidence to clarify their licensing scope and ensure accountability for any subsequent actions.
(v) The framework should facilitate communication between AI service providers and data owners~\cite{nguyen2023blockchain,wang2024sok}, enabling efficient attainment and documentation of licensing agreements. (vi) The framework should ensure the effective management and commercial sensitivity of licenses, safeguarding them against unauthorized access by third parties~\cite{chen2024catch}.

In this paper, we design, implement, and evaluate \textsc{IBis}, a blockchain-based framework for data and model copyright management, provenance, and lineage in AI model training processes. \textsc{IBis} empowers model owners to establish the provenance and lineage of their AI models and training datasets throughout retraining and fine-tuning processes, efficiently obtaining copyright licenses from the relevant copyright holders, and securely recording and renewing bilaterally signed copyright licenses as evidence of legal compliance. Our detailed contributions are as follows:

\begin{packeditemize}
    \item \textit{We propose a blockchain-integrated framework, \textsc{IBis}, to track data and model copyright management, provenance, and lineage.} \textsc{IBis} exhibits the following characteristics:
    \begin{itemize}
        \item[$\diamond$] \textit{Seamless integration} (addressing \textit{c-i}): By supporting iterative model retraining and fine-tuning, accommodating diverse copyright agreements through flexible license checks and renewals, and providing a unified API that integrates with existing contract lifecycle management software, the framework ensures minimal disruption to established model training and copyright management processes.

        \item[$\diamond$] \textit{Adaptability} (addressing \textit{c-ii and iii}): By establishing links between models in the model metadata, and integrating periodic license renewal checks via smart contracts, \textsc{IBis} supports ongoing model retraining and license renewal. Moreover, the on-chain license registry leverages blockchain's immutability property, allowing model owners and copyright holders  to retrieve their past licenses to prove regulatory compliance and avoid any disputes.

        \item[$\diamond$] \textit{Traceable registry} (addressing \textit{c-iv}): By deploying the on-chain, immutable registries for dataset metadata, licenses, and model metadata, the framework maintains authentic records of dataset and model relationships, ownership, and their copyright agreements. The bidirectional links between these records enables two-way traceability throughout data and model copyright management, provenance, and lineage processes.
          
        \item[$\diamond$] \textit{Blockchain-based multi-party signing} (addressing \textit{c-v}): By leveraging the identity management and digital signature capabilities offered by private-permissioned blockchains, \textsc{IBis} enables efficient and secure multi-party signing workflows between AI model owners and copyright holders, ensuring the establishment of legally compliant licensing agreements.
  
        \item[$\diamond$] \textit{Controllability} (addressing \textit{c-vi}): By implementing on-chain access control mechanisms and adhering to strict permission rules, \textsc{IBis} ensures that only authorized parties can access the information pertaining to training datasets, models, and licenses. Consequently, \textsc{IBis} facilitates an ecosystem encompassing many AI models, datasets, and licenses, enabling model and data owners to leverage the network effect of a unified platform while safeguarding their commercial sensitivity needs.
    \end{itemize}

    \item \textit{We implement a fully-functional prototype\footnote{Open released at \url{https://github.com/yilin-sai/ai-copyright-framework}.} based on the Daml smart contract language~\cite{bernauer2023daml} and Canton blockchain protocol~\cite{da2020canton}}. We adopted Daml and Canton's renowned privacy-preserving capabilities and modular design to implement a secure and commercial-sensitivity-preserving framework with six modules dedicated to license registration, management, and updating.

    \item \textit{We conduct a series of performance evaluations of \textsc{IBis}, especially its performance under a parameterized real-world scenario.} Evaluation results show that a model owner can retrieve a model's datasets and its licenses in approximately 1.5 and 3 seconds, respectively. This is irrespective of the number of model owners, datasets, and licenses hosted within the framework. Additionally, retrieving authorized models for a license takes approximately 1.5 seconds, regardless of the number of training datasets per model, model owners, and licenses within the framework. These results demonstrate scalability under varying numbers of users, datasets, models, and licenses.
    
\end{packeditemize}

The rest of the paper is organized as follows: Sec.\ref{sec-preli} provides background and related work. Sec.\ref{sec-framework} gives the system architecture and our design. The construction details of our framework, including data models and functional operations, are presented in Sec.\ref{sec-construction}. Sec.\ref{sec-imple} and \ref{sec-evaluation} present our implementation with performance evaluations.  Sec.\ref{sec-conclusion} offers conclusions and suggests avenues for future research.

\section{Technical Warm-ups}
\label{sec-preli}

\subsection{Background}
\label{sec-bkgd}

\begin{figure}[!htb]
  \centering
  \includegraphics[width=\linewidth]{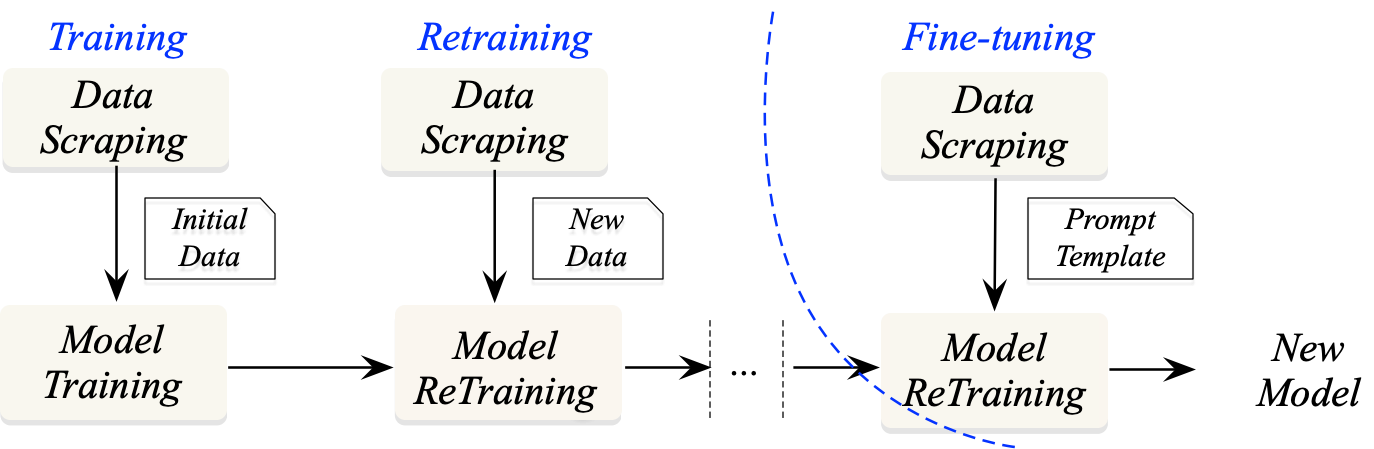}
  \caption{Typical AI model training process.}
  \label{fig:trainingprocess}
\end{figure}

\noindent\textbf{AI model training.} 
In general, the training process for AI models is continuous and iterative, containing \textit{training}, \textit{retraining}, and \textit{fine-tuning}~\cite{kreuzberger2023mlops}. 
As seen in Fig.\ref{fig:trainingprocess}, training begins with data collection, where initial training datasets are gathered through \textit{data scraping}. These datasets are then fed into the \textit{model training} step, where a preliminary model is trained. To ensure the model remains effective and up-to-date, it undergoes periodic retraining with newly collected data, allowing it to adapt to new information. Additionally, a model may undergo a \textit{fine-tuning} phase, where it is slightly retrained to meet specific domain requirements, enhancing its accuracy and relevance for targeted applications.

\vspace{-0.3em}

\smallskip
\noindent\textbf{Copyrights.} Copyright grants creators exclusive rights to their original expressions, such as literary, artistic, and musical works. This legal framework safeguards creators' rights, allowing them to control how their work is used, reproduced, and distributed.

\textit{Copyright protection} is automatic upon creation, but it is implicit, requiring additional steps for proper protection. First, registering the work with the copyright office offers authoritative legal evidence of ownership and eligibility for statutory damages in case of infringement.  Second, adding a copyright notice (©) with the creator's name and the year of creation informs others of the copyright claim~\cite{litman2016notice}, akin to a signature on a picture. Additionally, NFTs~\cite{wang2021non} offer a novel method for embedding ownership of digital art via blockchain, with ownership automatically claimed upon minting.

\textit{Licensing} is the primary method for granting or transferring the rights to a work. Creators can control the scope of rights by specifying terms and conditions within the license agreement. Licenses may vary widely, from granting broad permissions to restricting usage to specific purposes or timeframes.

\begin{table*}[t]
  \centering
  \caption{Comparison of \textsc{IBis} with related solutions}
  \label{tab:comparison}
  \resizebox{\linewidth}{!}{
  \begin{tabular}{lcccccc}
    \toprule
    \textbf{Solutions} & \textbf{\makecell{c-i: \\ Traceability}} & \textbf{\makecell{c-ii:\\ Immutability}} & \textbf{\makecell{c-iii: \\Multi-party Licensing}} & \textbf{\makecell{c-iv: \\Reuse Control}} & \textbf{\makecell{c-v: \\Lineage}} & \textbf{\makecell{c-vi:\\ Scalable Querying}} \\
    \midrule
    Generic NFTs~\cite{yu2024maximizing} & \checkmark & \checkmark & $\times$ & $\times$ & $\times$ & $\times$ \\
    Provenance DRM~\cite{savelyev2018copyright,jing2021blockchain,wang2022image} & \checkmark & \checkmark & \checkmark & $\times$ & $\times$ & $\times$ \\
    LicenseChain~\cite{borzunov2022distributed,gao2023gradientcoin,liu2024decentralized,yu2023ironforge} & \checkmark & \checkmark & \checkmark & $\sim$ & $\times$ & $\times$ \\
    \midrule
    \textsc{IBis} (this work) & \checkmark & \checkmark & \checkmark & \checkmark & \checkmark & \checkmark \\
    \bottomrule
  \end{tabular}
  }
\end{table*}

\subsection{Related Work}
\label{subsec-rw}

\noindent\textbf{Protecting copyright/data in AI.}  Data and copyright protection in AI services is a long-standing topic. Existing methods can be classified in several aspects~\cite{meurisch2021data}: Data-modifying approaches involve modifying or sanitizing user data to unlink them from specific individuals (e.g., k-anonymity~\cite{gedik2007protecting}, differential privacy~\cite{xu2019achieving}, and watermarking~\cite{zhang2018protecting}). This minimizes the risk of reidentification by removing or concealing Personally Identifiable Information (PII). Data-encrypting approaches encrypt user data to ensure integrity and confidentiality during data sharing, leveraging techniques such as homomorphic encryption~\cite{gilad2016cryptonets} and secure Multi-Party Computation (MPC)~\cite{rouhani2018deepsecure}. Data-minimizing approaches aim to boost efficiency by reducing the volume of personal data needed~\cite{shokri2015privacy}, often observed in general model training where PII data are not required during training and minimally during inference. Data-confining approaches involve AI methods that operate without sharing PII data beyond user boundaries~\cite{servia2018privacy}, ensuring data integrity and confidentiality while enabling effective personalization through local access to personal data.

\smallskip
\noindent\textbf{Blockchain-empowered copyright management.} 
Liang et al.~\cite{liang2020circuit} employed smart contracts to establish a homomorphic encryption mechanism aimed at safeguarding circuit copyrights.  Liu et al.~\cite{liu2024blockchain} employed a blockchain-based fraud-proof protocol to secure ownership rights over AIGC (artificial intelligence-generated content). Numerous similar solutions are outlined in studies such as~\cite{savelyev2018copyright,jing2021blockchain,wang2022image}.
It is worth noting that most existing blockchain-based studies treat each copyright merely as a form of non-fungible online property, akin to an NFT. However, this approach restricts its practical utility in real-world scenarios that require varied operations like registration, renewal, and termination -- features that our framework offers in contrast.

\smallskip
\noindent\textbf{Leveraging blockchain in AI.} Recent studies made efforts to empower AI and foundational models with blockchain technology, aiming to build a more robust and trustworthy AI in distributed environments. IronForge~\cite{yu2023ironforge} proposes a decentralized federated learning framework that integrates a distributed ledger and a Directed Acyclic Graph (DAG)-based data structure to asynchronously distribute training resources. Petals~\cite{borzunov2022distributed} is a distributed deep learning system that can effectively operate and refine complex models. It utilizes volunteer computing, outperforming traditional RAM offloading, particularly in autoregressive inference tasks.  BlockFUL~\cite{liu2024decentralized} introduces a decentralized federated unlearning framework that utilizes a redesigned blockchain structure leveraging Chameleon Hash. It decreases the computational and consensus costs associated with unlearning tasks. GradientCoin~\cite{gao2023gradientcoin} introduces a theoretical concept for a decentralized LLM that functions akin to a Bitcoin-like system.

\smallskip
\noindent\textbf{Comparison.} Table~\ref{tab:comparison} summarizes how \textsc{IBis} compares with representative blockchain-based copyright, provenance, and licensing systems. We align our comparison with the six core challenges (c-i to c-vi) outlined in the Introduction. Existing platforms often address immutability or traceability, but fall short in fine-grained license control, model retraining lineage, and scalable graph-based querying. 

Our design diverges from these existing models by opting not to rely on computationally heavy cryptographic primitives. Instead, we utilize blockchain-based multi-party signing and customizable access control mechanisms, which streamline the establishment and renewal of licensing agreements. This  ensures secure and transparent lineage tracking directly within AI model training workflows, offering a novel approach that enhances both compliance and operational efficiency without overhead associated with traditional cryptographic methods.

\section{Proposed Design: \textsc{IBis}}
\label{sec-framework}

\subsection{Design Overview}

\noindent\textbf{Roles.}
We distinguish between two pivotal roles: \textit{AI Model Owners} (AOs) and \textit{Copyright Owners} (COs). AOs act as representatives of the creators or uploaders of AI models, who construct, train, maintain, and commercialize the AI models. In our framework, their responsibilities include the categorization of data, acquirement of licenses, and registration of dataset/model metadata onto the blockchain. COs are the rightful copyright holders of training data with the authority to license their data, encompassing content creators and media companies among others. Their involvement extends to the drafting and bilateral signing of license agreements, ensuring regulatory compliance, and the protection of intellectual property rights. 
Likewise, a foundational model owner is also a CO from the point of view of an extended AO. In this scenario, datasets utilized in developing the foundational model are licensed by a separate set of COs, encompassing the licensing of derivative works. Distinguishing between these two CO roles is not imperative within \textsc{IBis}, as it can effectively keep track of complex data and model relationships (see Sec.\ref{sec-operation}).

\begin{figure}[!htb]
  \centering
  \includegraphics[width=\linewidth]{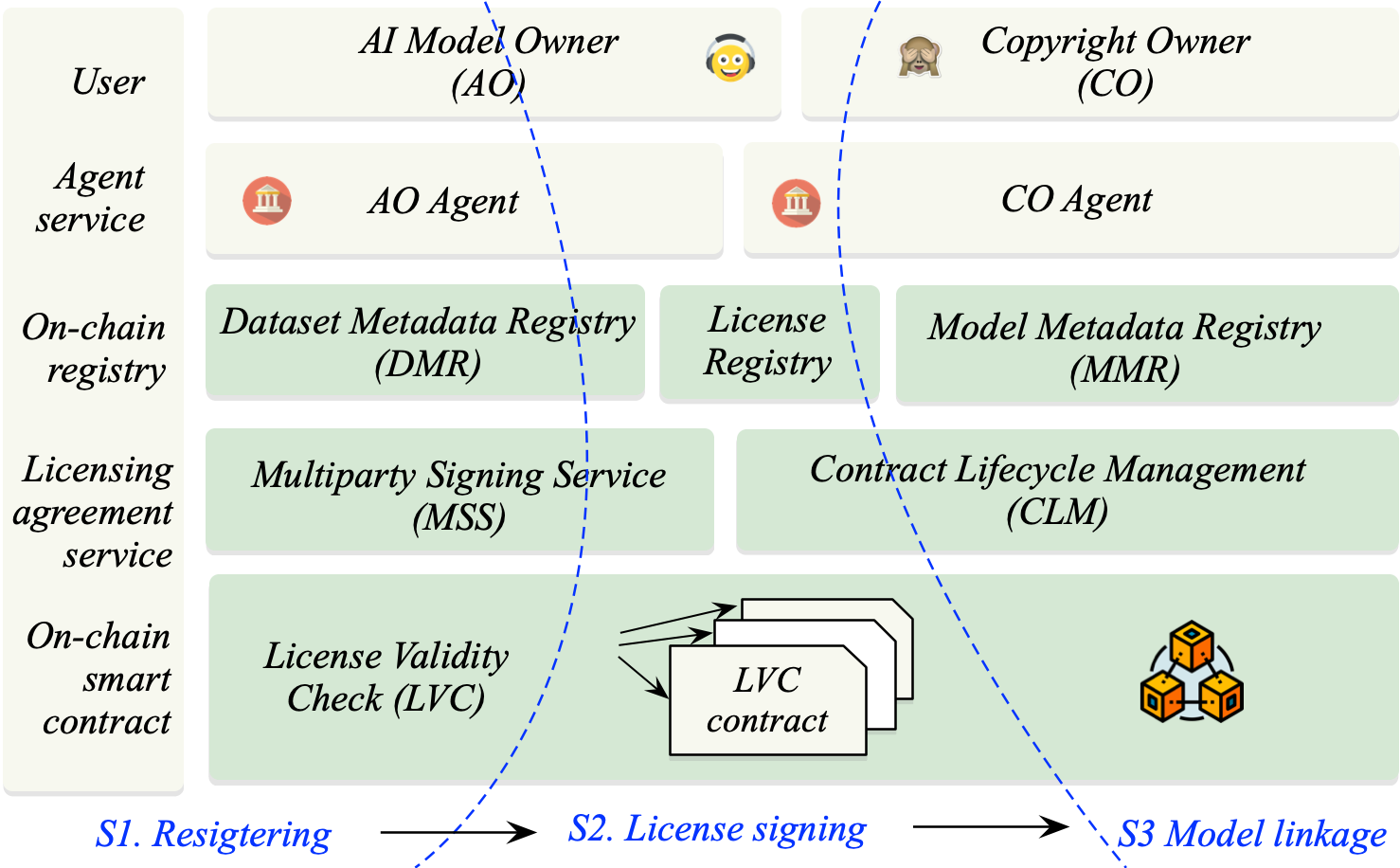}
  \caption{Architectural design.}
  \label{fig:sysarch}
\end{figure}

\smallskip
\noindent\textbf{Architecture.}
We envision an ecosystem that empowers both AOs and COs to harness the network effect of a unified platform to train and use numerous AI models and datasets. For example, CO could license the same dataset to multiple AOs and reap the benefits of a pay-as-you-go model for dataset usage or derived work within the same platform. Therefore, our proposed framework, \textsc{IBis} (cf. Fig.\ref{fig:sysarch}), is designed to integrate a blockchain network hosted by a subset of AOs and COs. While established and commercially significant AOs and COs may operate blockchain nodes, others only require the capability to connect to one of them via an agent.

At its core, the blockchain network serves as the backbone, facilitating secure and transparent interactions between AOs and COs. The system is architected to abstract the complexities of blockchain interaction through an agent service, offering user interface, authentication, and request buffering. The agent service acts as a bridge, connecting the AOs and COs with the blockchain, thereby enabling efficient metadata registering and licensing processes. Additionally, the system architecture includes dedicated components for handling dataset metadata registration, bilateral license signing, and model metadata management, each playing a crucial role in the overall workflow of AI model training and copyright handling.

\subsection{Key Modules} 

\textsc{IBis} has the following six main modules (green in Fig.\ref{fig:sysarch}):

\begin{itemize}
    \item \textit{Dataset Metadata Registry} (DMR) maintains an on-chain metadata record for each dataset scraped by AOs. These records include details such as the dataset's CO and its source URL.

    \item \textit{License Registry} records copyright licenses that are bilaterally signed by the corresponding CO and AO, serving as evidence of data use agreements. 
    When a dataset is licensed, a two-way linkage is established between the license record and the DMR record of the dataset. A single license may cover multiple datasets.

    \item \textit{Model Metadata Registry} (MMR) stores the metadata of a model once it has been trained. This metadata includes the model's identifier, as well as the identifiers of training datasets and source models. It maintains a persistent record of the datasets and source models utilized in models' training, thereby establishing data provenance.

    \item \textit{Multi-party Signing Services} (MSS) orchestrate communication between AOs and COs. It handles tasks such as sending license request emails to COs and returning license drafts to AOs. Most importantly, leveraging the identity management and digital signature capabilities of the blockchain, MSS ensures secure multi-party signing processes for establishing copyright licenses.

    \item \textit{Contract Lifecycle Management} (CLM) provides a unified API to interface with various external CLM software solutions that manage licenses. This approach ensures compatibility with a range of CLM software solutions, minimizing disruption to COs' existing workflows.

    \item \textit{License Validity Check} (LVC) employs smart contracts to verify the validity of a license based on a set of environment variables, including the current date, AO's operating location, and any other variables that could potentially contravene terms and conditions stipulated in the license. Our framework allows the creation of custom LVC smart contracts targeting different licence types.

\end{itemize}

Notably, both MSS and CLM are implemented as off-chain services. They are essential for facilitating interactions that precede and follow blockchain transactions, but they do not operate directly on the blockchain.

\subsection{Stages} 
For the initial model training, the workflow of our framework can be segmented into the following three stages:

\smallskip
\noindent\textit{\textbf{S1.} \underline{Dataset registering and license check:}} This involves dataset categorization, metadata registration, and license checks via smart contracts to ensure copyright compliance.

Specifically, the workflow begins with dataset categorization, where datasets are organized into specific categories based on their content, source, and usage. This categorization facilitates efficient retrieval and management of datasets throughout the AI model development process. Following categorization, metadata registration takes place via DMR, recording crucial details such as data descriptions, authorship information, and usage rights within a structured format. This step ensures that comprehensive information about the datasets is readily accessible and referenced during their lifecycle. Finally, license checks are conducted via LVC smart contracts, utilizing automated processes to verify the authenticity and compliance of licenses associated with the datasets. Smart contracts ensure that AI model training and usage adhere to copyright agreements, providing a streamlined and compliant workflow for managing datasets and their associated licenses.

\smallskip
\noindent\textit{\textbf{S2.} \underline{License drafting and bilateral signing:}} In case of failed license checks, this stage involves drafting and bilateral signing of licenses, facilitated by the MSS and CLM.

Upon identifying any missing, expired, or reworked licenses during stage S1, the process swiftly progresses to crafting bespoke license agreements tailored to the unique datasets and their intended applications. Leveraging recent advances in MSS technology, stakeholders embark on bilateral negotiations to refine the terms of these agreements, culminating in their formal agreement/contract through digital signatures. Finalization of agreements occurs only upon the attainment of signatures from both parties. Subsequently, CLM solutions seamlessly interface with current systems, streamlining contract management duties including drafting, approval workflows, and compliance oversight.

\smallskip
\noindent\textit{\textbf{S3.} \underline{Model registering and copyright owner notification:}} In this stage, post-training models are recorded in on-chain MMR, creating a bidirectional linkage between models and training datasets. Notifications may be dispatched to COs as stipulated by the licensing agreement.

\subsection{Details of Each Stage}
\begin{figure}[!htb]
  \centering
  \includegraphics[width=0.9\linewidth]{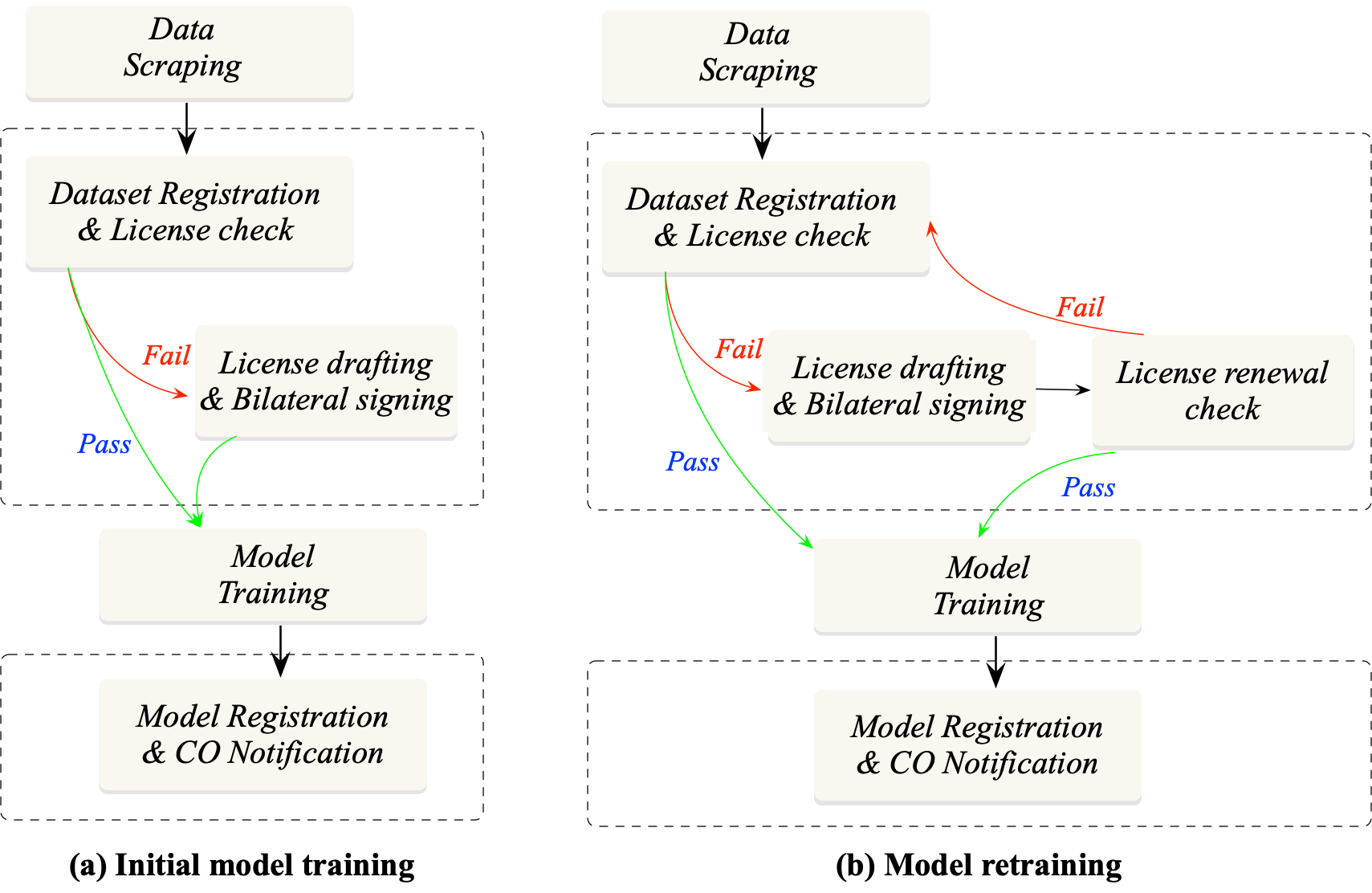}
  \caption{Integration with existing AI workflow.}
  \label{fig:initialtrainingflow}
\end{figure}

\begin{figure}[!htb]
  \centering
  \includegraphics[width=\linewidth]{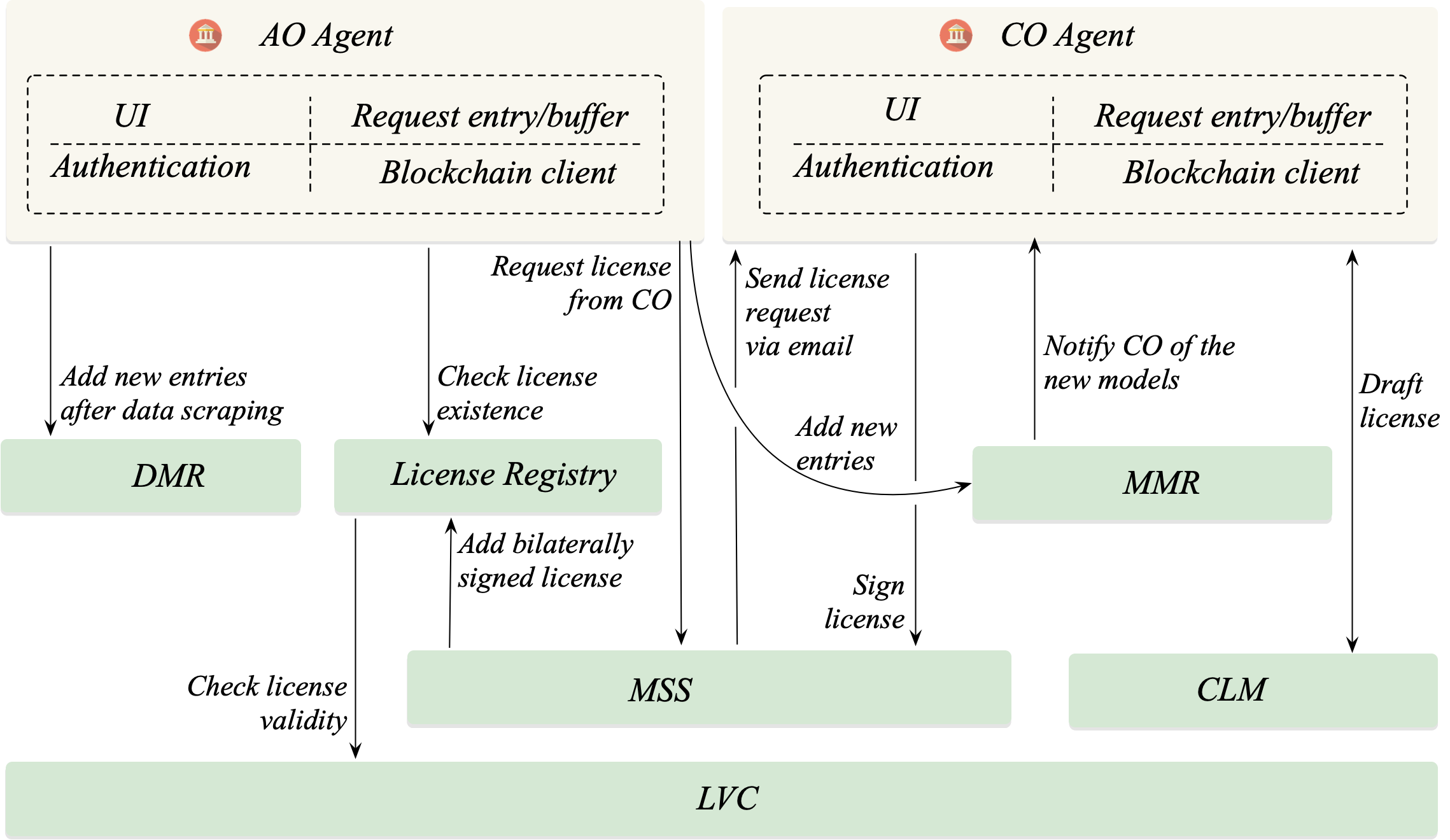}
  \caption{Functional design across each module.}
  \label{fig:sysdetail}
\end{figure}

The flowchart in Fig.\ref{fig:initialtrainingflow}.a illustrats how \textsc{IBis} is integrated into existing AI model training. Next, we discuss the main stages in detail, and Fig.\ref{fig:sysdetail} depicts interactions across \textsc{IBis} modules.

\subsubsection{\underline{Dataset metadata registering and license check}}
During or after the initial data scraping process, AOs categorize the data into one or more datasets and register the metadata of each dataset on the blockchain. The on-chain DMR maintains a metadata record for each dataset, containing details such as its CO\footnote{The method for acquiring copyright owner's information during data scraping is out of the scope of this paper.} and source URL (refer to Sec.\ref{sec-datamodel} for detailed information on data model specifics). Notably, \textsc{IBis} operates under the assumption that COs are uniquely identifiable, and the AO organizes the data in such a manner that each dataset is associated with only one copyright owner. A corner case arises when a dataset originates from the public domain and therefore lacks a distinct owner. In such instances, we stipulate the use of a designated copyright owner identifier, ``public-domain", to account for public domain data.

Datasets registered in DMR are not automatically eligible to become training data. To adhere to copyright laws, each dataset must undergo a vetting step, involving a check for the existence and validity of its license. During this step, AO extracts attributes of a dataset and queries the license registry on the blockchain for a corresponding license. 
Sec.\ref{sec-operation} describes the license registry search mechanism. 
Once a license is found, its validity is checked using the LVC smart contract. 
Only a dataset that passes the license check is deemed eligible for model training. 
Regarding the corner case of public domain data, the license registry is preloaded with a public domain license that always passes the LVC. Here, the mechanism of an LVC can vary depending on the type of license. For instance, certain licenses may impose geographic restrictions, while others may have time or number of use limits. Consequently, our framework facilitates the creation of different LVC smart contracts to accommodate such diverse and complex conditions. We define a generic LVC interface contract to dynamically determine which specific LVC contract to utilize based on the license being evaluated.

After the dataset successfully passes the license validity check, the licenseId attribute in its metadata will be updated to reference the valid license. Additionally, the dataset's identifier will be added to the license's datasetList attribute. This establishes a bidirectional linkage between a dataset and its corresponding license.

\subsubsection{\underline{License agreement drafting and bilateral signing}}
If the license existence or validity checks fail, AO must initiate a license agreement drafting and signing stage to obtain a license from the CO. This stage commences with AO sending a license request to the corresponding CO. This request is routed through MSS, which then generates an email containing the request details and along with a link for CO to take necessary actions. 
One of the actions involves drafting a license agreement based on the request. The CO executes this drafting action by invoking the API via CLM. Connectors that interface with various external CLM software solutions implement this API. This design is predicated on the understanding that COs often rely on proprietary software to draft their licensing agreements and manage data subscriptions. Thus, we leverage the CLM's API and connectors to ensure compatibility with a range of existing CLM software solutions, minimizing disruption to the copyright owners' existing workflow.

\begin{wrapfigure}{r}{4cm}
  \centering
  \includegraphics[width=\linewidth]{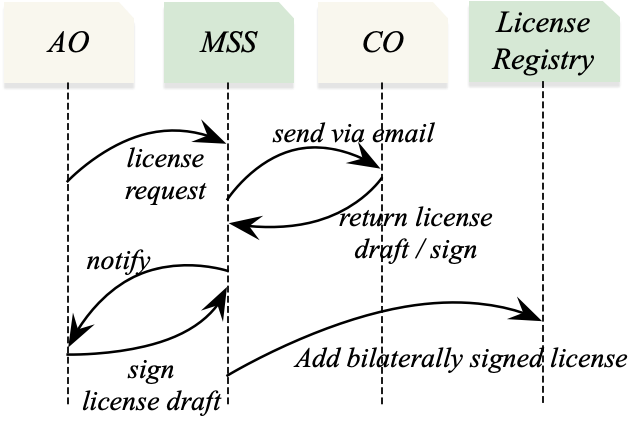}
  \caption{MSS workflow.}
  \label{fig:mps}
\vspace{-0.7em}
\end{wrapfigure}

Once the CO drafts the license agreement using the CLM software, the agreement is transmitted back to the framework via the CLM connector and API. Subsequently, CO and AO engage with the MSS to generate a bilaterally signed license agreement (cf. Fig.\ref{fig:mps}). The signed license agreement is then recorded in the license registry. The dataset's DMR record is also updated to reference the newly acquired license, thereby concluding the license agreement drafting and signing stage. Here, MSS utilizes an email list comprising the email addresses of AOs and COs. Email addresses can be added during user signup or input by the counterparty.

\subsubsection{\underline{Model registration and CO notification}}
The previous two steps empower the AO to accumulate a pool of licensed datasets that can be legitimately employed for AI model training. To establish data and model provenance, it is imperative to maintain a reliable record of the datasets and hyper-parameters utilized in each model's training. This is accomplished through the creation of MMR on the blockchain. Following the training of a model, its metadata, comprising its identifier, hyper-parameters, and the identifiers of the training datasets, are recorded on the MMR. 

Furthermore, for each training dataset, its DMR record is updated to include the identifier of the new model. This establishes a bidirectional linkage between a model and its training dataset. Finally, depending on the terms outlined in the licensing agreement, the framework may dispatch notifications to the respective copyright owners, informing them of the new model trained using their data.

\subsection{AI Model Retraining and Fine-tuning}

The system architecture outlined above not only supports initial model training but also facilitates model retraining and fine-tuning, wherein newly collected data are integrated into the model. As new data are scraped and collected, DMR continues to expand, while new licenses are acquired and stored in the license registry, ensuring the legitimacy of new datasets. Consequently, during the AI model retraining and fine-tuning, the initial two stages remain unchanged.

However, in retraining and fine-tuning scenarios, it is possible that the original model may need a license renewal at the time of retraining or fine-tuning, as one or more licenses of its training datasets might have become invalid (e.g., expired). Therefore, before retraining or fine-tuning the model, an additional stage is necessary to verify data eligibility by determining if the model requires a license renewal.
Fig.\ref{fig:initialtrainingflow}(b) depicts the flowchart of actions during model retraining or fine-tuning. Sec.\ref{sec-renewalcheck} presents the detailed mechanism used to conduct the license renewal check.

Moreover, compared to the stages in the initial training process, a divergence arises in the final stage concerning the establishment of data provenance. The new model is a culmination of the original model merged with new datasets. Hence, when recording a new entry in MMR for the retrained/fine-tuned model, in addition to recording the identifiers of the training datasets, the AO must also record the identifier of the original model. This facilitates data provenance and model lineage throughout the iterative model retraining or fine-tuning process. Meanwhile, the metadata of the original model must be updated to include a reference to the new model, thus establishing a bidirectional linkage between the new model and its source model. Complete information on model metadata specifics can be found in Sec.\ref{sec-datamodel}.

\section{IBIS with Detailed Workflow}
\label{sec-construction}


\subsection{Data Models} \label{sec-datamodel}

The main attributes that \textsc{IBis} framework supports can be broadly grouped as follows (see Table~\ref{tab:attributes}):

\begin{itemize}
    \item \textit{Dataset attributes}: A dataset is uniquely identified by a $\mathsf{datasetId}$ and sourced from a specific URL. It includes information on the copyright owner, associated license, and models trained on it. Additionally, the dataset's ownership and creator are tracked through the CO's $\mathsf{copyrightOwnerId}$.
    
    \item \textit{License attributes}: Each license has a distinct $\mathsf{licenseId}$ and encompasses a defined scope, typically a URL/URI. It includes details like copyright ownership, digital signatures of owners, and validity timestamps. The license type identifier $\mathsf{typeId}$ aids in determining the applicable LVC smart contract for license validation. Moreover, it lists the datasets covered under the terms of the license.
    
    \item \textit{Model attributes}: AI models are identified by a unique $\mathsf{modelId}$ and associated with owners. They utilize datasets for training, which are listed within the model's attributes. Retrained models reference a source $\mathsf{sourceModelId}$, and any subsequent models derived from it are listed as child models in $\mathsf{childModelList}$. This structure establishes the lineage of models and facilitates the tracking of relationships between models and data within AI services.
\end{itemize}

\begin{table}[t]
\caption{Dataset, license, and model attributes.}
  \label{tab:attributes}
  \resizebox{\linewidth}{!}{
  \begin{tabular}{cl p{7cm}}
    \toprule
    & \multicolumn{1}{c}{\textbf{Atttributes}} & \textbf{Descriptions} \\
    \midrule
    \multirow{6}{*}{\rotatebox{90}{\textbf{Dataset}}} & $\mathsf{datasetId}$ & Unique identifier of the dataset. \\
    & $\mathsf{sourceUrl}$ & URL from which the dataset was scraped. \\
    & $\mathsf{copyrightOwnerId}$ & Unique identifier of the copyright owner. \\
    & $\mathsf{licenseId}$ & Unique identifier of the copyright license. \\
    & $\mathsf{modelList}$ & List of unique identifiers of the models trained on this dataset. \\
    & $\mathsf{modelOwnerId}$  & Unique identifier of AO who scrapes and adds this dataset.\\
    
    \cmidrule{1-3}

    \multirow{12}{*}{\rotatebox{90}{\textbf{License}}} & $\mathsf{licenseId}$ & Unique identifier of the dataset. \\
    & $\mathsf{scope}$ & URL to dataset. Datasets pointed by this URL fall within the scope of the license. \\
    & $\mathsf{copyrightOwnerId}$ & Unique identifier of the copyright owner. \\
    & $\mathsf{copyrightOwnerSignature}$ & Digital signature of the copyright owner. \\
    & $\mathsf{modelOwnerId}$  & Unique identifier of the model owner.\\
    & $\mathsf{modelOwnerSignature}$  &  Digital signature of the model owner. \\
    & $\mathsf{validFrom}$  &  Timestamp indicating when the license takes effect. \\
    & $\mathsf{typeId}$  & Unique identifier of the license type. Used to determine the smart contract to check the license validity. \\
    & $\mathsf{datasetList}$ & List of identifiers of the datasets covered by this license. \\ 
    
    \cmidrule{1-3}

     \multirow{6}{*}{\rotatebox{90}{\textbf{Model}}} & $\mathsf{modelId}$ & Unique identifier of the AI model. \\
    & $\mathsf{modelOwnerId}$ & Unique identifier of the model owner. \\
    & $\mathsf{datasetList}$ &  List of unique identifiers of training datasets. \\
    & $\mathsf{sourceModelId}$ &  Unique identifier of the source model that has been retrained.  \\
    & $\mathsf{childModelList}$  & List of identifiers of the models trained based on this model. \\

  \bottomrule
  
\end{tabular}
}

\end{table}

We highlight two aspects. First, the data model is extensible, enabling AOs and COs to incorporate additional custom attributes as needed. For example, a storage URL can be included in dataset 
\begin{wrapfigure}{r}{4cm}
  \centering
  \includegraphics[width=0.9\linewidth]{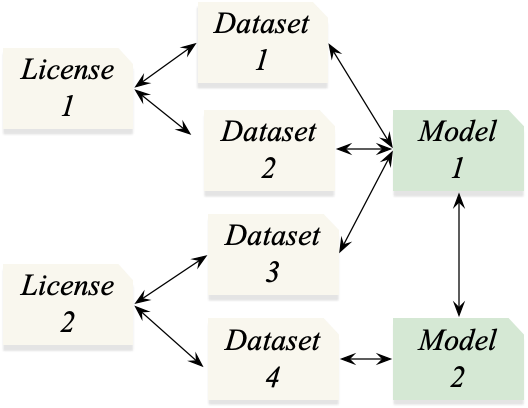}
  \caption{Model, dataset, and license relationships.}
  \label{fig:datamodelrel}
\vspace{-0.5em}
\end{wrapfigure}
metadata to indicate where AO stores the dataset. A license can include custom attributes such as expiration date, exclusivity, and other terms and conditions. Second, a web path is employed to delineate the scope of what is being licensed, considering that the majority of AI models are trained using online data.

\smallskip
\noindent\textbf{A running example.} Fig.\ref{fig:datamodelrel} illustrates the logical interrelation among the three data models, within an example scenario where Model-1 is initially trained using three datasets, and subsequently retrained with a fourth dataset to yield Model-2. The two models are linked through Model-2's $\mathsf{sourceModelId}$ attribute and Model-1's $\mathsf{childModelList}$.
A dataset and a model are linked through the model's $\mathsf{datasetList}$ and the dataset's $\mathsf{modelList}$. A license and a dataset are linked through the license's $\mathsf{datasetList}$ and the dataset's $\mathsf{licenseId}$.

\smallskip
\noindent\textbf{Integrity binding with off-chain data.}
While the actual datasets and models are stored off-chain, \textsc{IBis} ensures integrity through the use of cryptographic hashes. Each dataset and model is associated with a content-addressable identifier, typically computed using a collision-resistant hash function (e.g., SHA-256). These hashes are recorded on-chain as part of the metadata entries in the DMR and MMR smart contracts. As a result, any tampering with off-chain content can be detected by recomputing and comparing the hash values.

Optionally, the system may integrate with decentralized storage platforms (e.g., IPFS), where the content hash also serves as the retrieval address. This dual use of hashes enhances verifiability and offers a robust mechanism for ensuring data provenance and integrity even in off-chain storage.

\subsection{Obtaining Licenses} 
\label{sec-operation}

\noindent\textbf{Obtain dataset licenses.} 
The $\mathsf{getDatasetLicense}$ operation retrieves copyright license of a given a dataset identifier $\mathsf{datasetId}$. It initially searches the DMR hash map using the dataset $\mathsf{datasetId}$ as the key, which incurs a time complexity of $O(1)$. Depending on whether the returned DMR record contains a $\mathsf{lisenceId}$, this operation either retrieves the license by $\mathsf{licenseId}$ or searches for a relevant license in the license registry as follows: 

\begin{itemize}

    \item \textit{Retrieve the license with $\mathsf{licenseId}$}: This operation queries the license registry hash map using the $\mathsf{licenseId}$ as the key. As the resulting time complexity is $O(1)$, the overall time complexity remains the same.

    \item \textit{Search for a relevant license}: This operation extracts the dataset's $\mathsf{copyrightOwnerId}$ and performs a search on the license registry using $\mathsf{copyrightOwnerId}$, which involves a complexity of $O(1)$. This search may yield a list of licenses with the same $\mathsf{copyrightOwnerId}$ (albeit with different scopes).
    Finally, a scan is conducted on the list of licenses to filter out the licenses with irrelevant scopes. Our framework operates on the premise that license scopes do not intersect, ensuring each dataset corresponds to at most one license. As CO is unlikely to have many licenses with the same AO for practical reasons, one can assume the size of this list to be small. 

\end{itemize}

\smallskip
\noindent\textbf{Obtain model licenses.} 
Given a model identifier $ \mathsf{modelId} $, the $ \mathsf{getModelLicenses} $ operation retrieves the licenses of its training datasets. This operation begins by identifying the training datasets associated with the provided model and its upstream source models. Each dataset is linked to a single license, and the operation retrieves these licenses by traversing the graph where models, datasets, and licenses are nodes. The graph includes edges representing model-to-dataset, dataset-to-license, and model-to-model relationships (as depicted in Fig.\ref{fig:datamodelrel}).
For a specific model, the operation traverses $ T $ datasets linked to it, retrieves their associated licenses, and deduplicates them if necessary. The time complexity for this operation is $ O(T) $ per model, as retrieving the licenses for $ T $ datasets involves processing $ T $ edges.
When the operation considers all $ M $ models, the total complexity becomes $ O(|M||T|) $, where $ T $ is the number of datasets linked to each model. This complexity accounts for accessing training datasets, retrieving licenses, and aggregating results. Since $ T $ is typically much smaller than $ D $ (the total number of datasets in the graph), the traversal focuses only on the relevant portion of the graph. Therefore, the overall complexity scales linearly with the number of models and their associated datasets.

\subsection{Validating Licenses} 

\noindent\textbf{Check license validity.} 
Given license data and environment variables as transaction inputs, the $\mathsf{checkLicenseValidity}$ operation verifies the validity of the license. Environment variables include the current date, the operating locations of AOs, and other variables that could potentially contravene the terms and conditions stipulated in the license agreement. 

First, we need to locate the corresponding LVC smart contract for validating the license. This can be accomplished using another hash map where the license type $\mathsf{typeID}$ serves as the key and LVC's address as the value. Consequently, this lookup operation can be performed in constant time, i.e., $O(1)$. Next, we need to invoke the identified LVC contract to determine the license validity. 
The time complexity of the LVC contract is directly proportional to the number of environment variables that need validation. We abstract this time complexity as $O(|K|)$, where $\textbf{K}$ is the set of environment variables to validate. The overall time complexity is $O(1) + O(|K|)=O(|K|)$.

Here, we give an example (Listing~\ref{lst:lvc-timebound}). This LVC template is triggered by the $\mathsf{checkLicenseValidity}$ operation and verifies that the license is currently active based on temporal conditions. The modular design of LVCs allows for extensible contract templates supporting various types of constraints (e.g., geographic, exclusivity, usage count), which can be selected based on the license's $\mathsf{typeId}$.

\begin{lstlisting}[style=JavaScript, caption={Example LVC smart contract template for validating a time-bound license.},label={lst:lvc-timebound},basicstyle=\ttfamily\scriptsize,xleftmargin=.05\columnwidth, xrightmargin=.05\columnwidth]
template TimeBoundLVC with
    licenseId: Text
    validFrom: Time
    validUntil: Optional Time
    currentTime: Time
  where
    signatory licenseVerifier

    choice CheckValidity : Bool
      controller licenseVerifier
      do
        let isValidStart = currentTime >= validFrom
        let isValidEnd = case validUntil of
                            Some end -> currentTime <= end
                            None -> True
        return (isValidStart && isValidEnd)
\end{lstlisting}

\subsection{Obtaining Data and Models via Licenses}

\noindent\textbf{Obtain licensed datasets.} 
The $\mathsf{getLicensedDatasets}$ operation retrieves the list of dataset identifiers $\mathsf{datasetId}$s each with a valid license. It entails executing $\mathsf{getDatasetLicense}$ and $\mathsf{checkLicenseValidity}$ operations for each dataset. Given $\textbf{D}$ datasets, the overall time complexity is $O(|D| \times \{O(1)+O(K)\}=O(|D||K|)$.

\smallskip
\noindent\textbf{Obtain authorized datasets by license.}
Given a license identifier $\mathsf{licenseId}$, the $\mathsf{getDatasetsByLicense}$ operation retrieves datasets covered by the license. This operation performs a search of the DMR using the $\mathsf{licenseId}$, resulting in a time complexity of $O(1)$.

\smallskip
\noindent\textbf{Obtain authorized models by license.}  
Given a license identifier $ \mathsf{licenseId} $, the $ \mathsf{getModelsByLicense} $ operation retrieves the metadata of the models covered by the license. This operation begins by executing $ \mathsf{getDatasetsByLicense} $, which identifies all datasets associated with the given license. For each of these datasets, the operation performs a graph traversal to retrieve the models directly trained on them and any child models indirectly connected through shared datasets.  
The overall time complexity consists of two components: $ O(|D|) $ for identifying the datasets linked to the license and $ O(|M||T|) $ for traversing the graph to retrieve the connected models and their relationships. Consequently, the total complexity is $ O(D + |M||T|) $, where $ D $ is the total number of datasets, $ M $ is the total number of models, and $ T $ is the average number of datasets per model.

\smallskip
\noindent\textbf{Obtain model datasets.}
Given a model identifier $\mathsf{modelId}$, the $\mathsf{getModelDatasets}$ operation retrieves the identifiers of its training datasets. During the first traversal or initialization, this operation iterates through the provided model and its upstream source models, retrieving the associated datasets. If the model has a set of $\textbf{M}$ upstream models, each linked to $\textbf{T}$ datasets, the time complexity is $O(|M| |T|)$ for the initialization phase.

\subsection{Renewing License}

License validity check is an ongoing task because a valid license may become invalid under certain circumstances (e.g., revoked or expired), necessitating AOs and COs to take appropriate actions to ensure continuous compliance with copyright laws. Following delineates how the framework facilitates license renewal checks and renewals.

\smallskip
\noindent\textbf{License renewal check.} \label{sec-renewalcheck}
The framework supports three types of license renewal checks (LRCs): license-driven, dataset-driven, and model-driven.

In license-driven LRC, AOs or COs conduct a periodic scan of the license registry, performing $\mathsf{checkLicenseValidity}$ on each license. If a license fails the validity check, an AO can execute the $\mathsf{getModelsByLicense}$ operation to gather the identifiers of datasets and models that depend on the invalid license. These can be added to a blacklist to prevent the use of those datasets and models in future training of new models or retaining. The specifics of how the blacklist is stored and managed fall beyond the scope of this paper.

Dataset-driven and model-driven LRC can be conducted on-demand before training a new model. In dataset-driven LRC, an AO can execute $\mathsf{getDatasetLicense}$ operation followed by the $\mathsf{checkLicenseValidity}$ operation for each training dataset to identify any dataset needing a license renewal. In contrast, in model-driven LRC, an AO can execute $\mathsf{getModelLicenses}$ operation followed by $\mathsf{checkLicenseValidity}$ operation for each license of the model, determining whether the model needs a license renewal.

\smallskip
\noindent\textbf{License renewal.} A license renewal involves adding a new bilaterally signed license to the license registry, rather than updating existing records. This enables AOs and COs to access all historical licenses to prove regulatory compliance and avoid any disputes. After a new license has been added, an AO can execute the $\mathsf{getModelsByLicense}$ operation to gather the identifiers of datasets and models that depend on the renewed license. Then the DMR records of datasets are updated to reference the new license. As the list of dependent datasets and models can now be considered eligible for training, their identifiers are also removed from the blacklist.

\subsection{More Operational Features}

\noindent\textbf{Operation atomicity.} It is observed that several stages (i.e., \textit{S.1}, \textit{S.3}, and License Renewal) involve the update of multiple records. Apart from ensuring integrity and immutability, another advantage of maintaining DMR, license registry, and MMR on-chain is that such multiple updates are guaranteed to be atomic. This is because smart contracts can ensure atomicity where the actions included in one transaction either all take effect or none of them take effect. Therefore, care needs to be taken during implementation to ensure that all updates within a stage should be included in the same transaction.

\smallskip \noindent\textbf{License linkage.}
In \textsc{IBis}, each dataset is assigned a license at the time of registration through the $\mathsf{licenseId}$ attribute. This linkage is immutable and ensures that all subsequent uses of the dataset are bound by the same licensing terms. The license itself enumerates the covered datasets via its $\mathsf{datasetList}$ attribute. This bidirectional mapping supports consistent enforcement of usage policies and enables automatic validation of licensing conditions during model training and retraining. The use of type-specific LVC contracts (see Sec.~\ref{sec-operation}) ensures that the correct license validation logic is applied in each case.

\smallskip \noindent\textbf{Dataset reuse.}
When a dataset is reused across multiple models, original licensing terms remain applicable. The reuse is authorized as long as the license associated with the dataset remains valid. During model creation or retraining, the system verifies license validity through $\mathsf{checkLicenseValidity}$ and ensures that all datasets listed in the model's $\mathsf{datasetList}$ are covered by licenses returned by $\mathsf{getModelLicenses}$. Since licenses may impose usage constraints (e.g., valid period, geographic restrictions), this step guarantees ongoing legal compliance for every instance of dataset reuse.

\smallskip \noindent\textbf{Traceability.}
To support auditing and provenance, \textsc{IBis} explicitly records the lineage of models and datasets. Each dataset maintains a $\mathsf{modelList}$ tracking all models trained on it, while each model retains a $\mathsf{datasetList}$ identifying its training inputs. For retrained models, the $\mathsf{sourceModelId}$ and $\mathsf{childModelList}$ attributes form a directed acyclic graph (DAG), capturing the complete history of derivations. These structural links are embedded in the smart contract state and are queryable via the $\mathsf{getModelDatasets}$ and $\mathsf{getModelsByLicense}$ operations. As a result, \textsc{IBis} provides fine-grained, tamper-proof traceability across both horizontal (multi-model) and vertical (model-chain) reuse scenarios.

\section{Implementation}
\label{sec-imple}

\subsection{Overview}
\label{sec-impl-overview}

\noindent\textbf{Prototype stack.}  
We deploy Canton~\cite{da2020canton}
network (v2.8.3) with three controlled nodes on AWS $\mathsf{t2.xlarge}$ instances (configuration by 4vCPU, 16GB RAM, Ubuntu22.04).  
Each node runs a \emph{Participant+Synchroniser} pair backed by
Docker-ised PostgreSQL15.

\smallskip
\noindent\textbf{Component map.}  
Fig.\ref{fig:uml} links the on-chain contracts
(\textsc{DMR}, \textsc{MMR}, \textsc{License}) to off-chain services
(MSS, CLM); Table~\ref{tab:resource} lists party-centric access scopes
enforced by Canton.

\smallskip
\noindent\textbf{Code footprint.}  
Core Daml templates $\approx380$ LOC, HTTP/gRPC glue $\approx210$ LOC, evaluation harness $\approx150$ LOC.

\smallskip
\noindent\textbf{Reproducibility.}  
We wrap up experiments and a single command line (i.e., \$\,$\mathsf{docker-compose up}$) can instantiate the full network and re-run Experiments 1–4; scripts are open-sourced at  
\url{https://github.com/yilin-sai/ai-copyright-framework}.

\subsection{Blockchain Platform}
\label{sec-blockchain}

We implement \textsc{IBis} in \textbf{Daml}
(Digital Asset Modelling Language) atop the
\textbf{Canton} ledger protocol.

\smallskip
\noindent\textbf{Contract-as-data model.}  
Daml templates codify both schema and behaviour. We can map
the Dataset/Model/License records of
Table~\ref{tab:attributes} to verifiable contracts with one-to-one
clarity.

\smallskip
\noindent\textbf{Ledger-level privacy.}  
Canton extends Daml’s role system
(\textit{signatory}, \textit{controller}, \textit{observer},
\textit{choice-observer}) so each node synchronises \emph{only} those
contracts it is authorised to see; commercial terms inside a
\textsc{License} stay invisible to third parties even on-ledger.

\begin{itemize}
\item \textit{observer}: A party that is guaranteed to see actions that create and archive the contract.
\item \textit{controller}: A party that can exercise a particular choice (i.e., invoke a function) on the contract.
\item \textit{choice observer}: A party that is guaranteed to see a particular choice being exercised on the contract.
\end{itemize}

\smallskip
\noindent\textbf{Why not Fabric / generic EVM?}  
Hyperledger Fabric chaincode can enforce ACLs but still replicates the
entire channel ledger to all members \cite{Androulaki_2018}.  
Public-chain EVM contracts disclose everything.  
Canton’s party-centric sync therefore offers the best trade-off between immutability and confidentiality for multi-tenant AI marketplaces.

Therefore, by adopting Daml and Canton, we facilitate the integration of many AOs and COs onto the same \textsc{IBis} platform, while ensuring that commercially sensitive license agreements between an AO and COs, dataset metadata, and model metadata remain concealed from other parties, even at the ledger level. 
Table~\ref{tab:resource} lists the resources in \textsc{IBis} along with their authorizations.

\begin{table}[t]
  \caption{Resources and authorisations.}
  \label{tab:resource}
  \centering
  \resizebox{\linewidth}{!}{
  \begin{tabular}{ccl}
    \toprule
    \textbf{Resource} & \textbf{Authorised Actor} & \textbf{Access Scope}\\
    \midrule
    DMR               & AOs       & Datasets added by the actor\\
    License Registry  & AOs/COs   & Licenses signed by the actor\\
    MMR               & AOs       & Models owned by the actor\\
    LVC API           & AOs/COs   & Licenses signed by the actor\\
    CLM API           & COs       & Licenses drafted by the actor\\
    \bottomrule
  \end{tabular}}
\end{table}

\subsection{Smart-Contract Implementation}
\label{sec-contract}

Listing~\ref{lst:daml-license} illustrates the license smart contract template, constructed  based on the data schema in Table~\ref{tab:attributes}. 
The $\mathsf{copyrightOwner}$ and
$\mathsf{modelOwner}$
are designated as signatories of the template (Line 10). This reflects the bilateral nature of the license agreement. Consequently, authorization from both parties is required to create a license contract. Subsequently, once the contract is created, both parties can access it, while no other party has access to this contract either on-chain or on-ledger. Hence, this implementation closely aligns with the access control requirements specified in Table~\ref{tab:resource}. In addition, the identifier attribute of each license serves as the primary key (specified by Line 12-13), which can facilitate efficient queries during the graph traversal. \label{sec-opt}

\begin{lstlisting}[style=JavaScript, caption={Daml smart contract template for license.},label={lst:daml-license},basicstyle=\ttfamily\scriptsize,xleftmargin=.05\columnwidth, xrightmargin=.05\columnwidth]
template License with
    licenseId: Text
    scope: Text
    copyrightOwner: Party
    modelOwner: Party
    validFrom: Time
    typeId: Text
    datasetList: [Text]
  where
    signatory copyrightOwner, modelOwner

    key (modelOwner, licenseId) : (Party, Text)
    maintainer key._1
\end{lstlisting}

Similarly, dataset and model metadata are represented in smart contract templates according to their corresponding data models, except that only the $\mathsf{modelOwner}$ is designated as the signatory of the $\mathsf{DatasetMeta}$ and $\mathsf{ModelMeta}$ contracts. This setup enforces two access control rules: 
\begin{itemize}
\item Only $\mathsf{modelOwner}$ has the authority to create and access these contracts. 
\item $\mathsf{modelOwner}$ is restricted to accessing $\mathsf{DatasetMeta}$ or $\mathsf{ModelMeta}$ contracts created by themselves.
\end{itemize}

\begin{figure}[!h]
  \centering
  \includegraphics[width=0.8\linewidth]{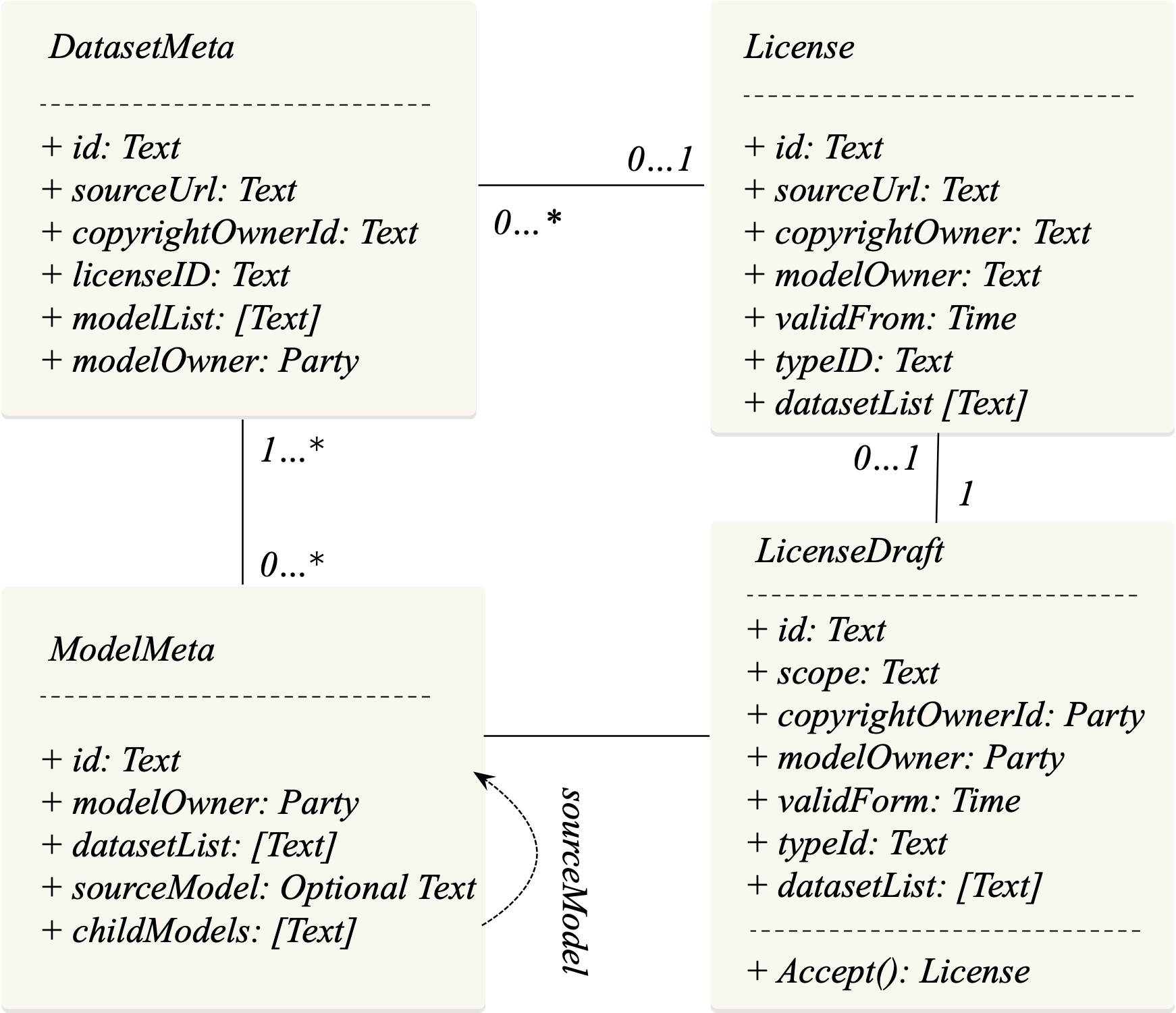}
  \caption{The class diagram of \textsc{IBis} implementation.}
  \label{fig:uml}
\end{figure}

Full implementation details are open sourced\footnote{Open released at \url{https://github.com/yilin-sai/ai-copyright-framework}.}. Fig.\ref{fig:uml} illustrates the class diagram of \textsc{IBis} design.

\subsection{Multi-party Signing}
The generation of a bilaterally signed license contract adheres to Daml's \textit{Propose and Accept Pattern}\footnote{{\url{https://docs.daml.com/daml/patterns/initaccept.html\#the-propose-and-accept-pattern}.}}. In this pattern, one party initiates a proposal contract, which the counterparty can either accept or reject. The acceptance (or rejection) is implemented as a contract choice\footnote{\url{https://docs.daml.com/daml/intro/4\_Transformations.html\#choices-as-methods}}, allowing the counterparty to exercise their decision. Upon exercising the accept choice, a result contract is generated, symbolizing the agreement between the two parties.

In our implementation, the proposal contract takes the form of the draft license agreement, and it is depicted in Listing~\ref{lst:daml-LicenseAggrement}. $\mathsf{LicenseAgreement}$ shares the same data schema as $\mathsf{License}$ (see Listing~\ref{lst:daml-license}), with the $\mathsf{copyrightOwner}$ designated as the signatory. This ensure only the $\mathsf{copyrightOwner}$ has the authority to create a contract. Alternatively, $\mathsf{modelOwner}$, as the controller of the $\mathsf{Accept}$ choice in Line 16, has the authority to exercise the choice, resulting in the creation of a $\mathsf{License}$ contract. This setup ensures that only the $\mathsf{modelOwner}$ specified in a $\mathsf{LicenseAgreement}$ contract has the authority to exercise the $\mathsf{Accept}$ choice of that contract.

\begin{lstlisting}[style=JavaScript, caption={Daml smart contract template for license agreement.},label={lst:daml-LicenseAggrement},basicstyle=\ttfamily\scriptsize,xleftmargin=.05\columnwidth, xrightmargin=.05\columnwidth]
template LicenseAgreement with
    id: Text
    scope: Text
    copyrightOwner: Party
    modelOwner: Party
    validFrom: Time
    typeId: Text
    datasetList: [Text]
  where
    signatory copyrightOwner

    key (copyrightOwner, id) : (Party, Text)
    maintainer key._1

    choice Accept: ContractId License
      controller modelOwner
      do create License
          with licenseId; scope; copyrightOwner; modelOwner; validFrom; typeId; datasetList
\end{lstlisting}

\subsection{Operational Footprint}
\label{sec-operational}

\noindent\textbf{Gas / storage.}  
A complete write–read cycle touches four contracts and stores
$\approx3.5$kB; even with $10^{3}$ datasets/models/licenses the ledger is
$\leq$ 15MB.

\smallskip
\noindent\textbf{Performance hooks.}  
Primary-key indices and batched Canton commits keep the
$P95$ end-to-end latency below 1.2 s at 1 000 req/s
(§\ref{sec-evaluation}) without heavyweight zk-proofs or HE / sMPC.

\section{Evaluation}
\label{sec-evaluation}

Our proof-of-concept \textsc{IBis} implementation was deployed on a private Canton blockchain comprising three nodes. All nodes were hosted on the same AWS EC2 $\mathsf{t2.xlarge}$ instance with four virtual CPUs and 16GB of RAM. While Daml provides a range of options for data storage, PostgresSQL, running within Docker containers, is chosen as the data storage to persist node data. Our source code of the performance test is available\footnote{Testing script: \url{https://github.com/yilin-sai/ai-copyright-framework}}. 
In the following sections, we present four sets of experiments: (1) baseline evaluation of core search operations, (2) comparative analysis of execution overhead against other blockchain-based designs, (3) scalability under varying load and storage conditions, and (4) robustness against adversarial attacks targeting provenance and signature integrity.

\subsection{Experiment-1: Searching Operations}
Our evaluation mainly focuses on three operations involving graph traversals: fetching model licenses using $\mathsf{getModelLicenses}$, model datasets using $\mathsf{getModelDatasets}$, and authorized models using $\mathsf{getModelsByLicense}$. The former two operations pertain to copyright management, while the latter concerns data provenance. To enhance the accuracy of performance testing, for every parameter configuration, the operation of $\mathsf{getModelLicenses}$ is executed ten times on ten randomly chosen models, with the average execution time and standard deviation calculated thereafter. Similarly, $\mathsf{getModelDatasets}$ undergoes execution on ten randomly selected models. As for $\mathsf{getModelsByLicense}$, the operation is performed on ten randomly chosen licenses, with the resultant average execution time and standard deviation recorded.

\smallskip
\noindent\textbf{Benchmarks and baselines.}  
In the absence of a unified benchmark for evaluating methods in this context, we focus on execution time as the primary evaluation criterion. This metric is reasonable and sufficient given that our method, \textsc{IBis}, is implemented on a blockchain and emphasizes efficient searching when providing copyright protection for models and datasets. Moreover, the focus on execution time aligns with common practices in evaluating non-functional requirements, especially for blockchain-based software platforms~\cite{LU2019564,10.1145/3408309, 8789542}, as functional requirements can vary significantly and are often implemented via heterogeneous protocol designs, making precise cross-method evaluations challenging.

To highlight the advantages of \textsc{IBis}, we compare it against two widely used architectural patterns: \textit{Flat/Sequential Search} and \textit{Full Graph Search}. These baselines represent common designs in real-world systems such as e-Commerce~\cite{10.1145/3408309} and e-Healthcare~\cite{10.1007/978-3-030-75078-7_60}, and also subsume recent blockchain-based copyright systems~\cite{r2-recent-1,r2-recent-2,r2-recent-3}.

\begin{itemize}
\item \textit{Flat/Sequential Search~\cite{10.1145/3408309,r2-recent-1,r2-recent-2}:} Assets are treated as isolated records with no structured relationships. Systems of this type store either individual entries (e.g., CIDs and ownership metadata~\cite{r2-recent-1}) or simple one-way chains per asset~\cite{r2-recent-2}. Each query scans a linear list or full dataset to validate, retrieve, or compare records. For example, $\mathsf{getDatasetLicense}$ requires $O(|D|)$ time compared to $O(1)$ in \textsc{IBis}; $\mathsf{getModelLicenses}$ is $O(|D||M|)$ vs.\ $O(|M||T|)$, and $\mathsf{getModelsByLicense}$ is $O(|D||M|)$ vs.\ $O(|D| + |M||T|)$. These access costs grow rapidly with scale and offer no built-in mechanism for capturing cross-asset dependencies.

\item \textit{Full Graph Search~\cite{10.1007/978-3-030-75078-7_60, r2-recent-3}:} More expressive systems represent licenses, evidence, and ownership as interconnected graph structures. For instance,~\cite{r2-recent-3} uses RDF triples and semantic links to connect works, claims, and complaints, with queries executed via external graph indexers. However, without pruning or indexing, operations such as $\mathsf{getModelLicenses}$ must walk all linked paths, resulting in $O(|M||T| + |D|)$  complexity compared to $O(|M||T|)$ in \textsc{IBis}. These systems capture richer semantics but suffer from high traversal costs and limited performance tuning.

\end{itemize}

\textsc{IBis} generalizes and improves upon both categories. It supports rich inter-asset semantics like the graph-based approaches while maintaining low-complexity queries through DAG indexing and caching. Because our design operates at the indexing layer, it can be added atop existing ledger infrastructures without changing their consensus or storage logic. This allows flat or graph-based systems to adopt \textsc{IBis} for efficient, provenance-aware queries and recursive license value flows.

\noindent\textbf{Experimental parameters.}
In real-world scenarios, the framework hosts data, including dataset metadata, license, and model metadata, contributed by various AOs and COs. These stakeholders engage in executing functional operations (outlined in Table~\ref{tab:operations}) to realize data provenance and copyright management. Ensuring the efficiency of operations, particularly those involving complex graph traversals, is paramount in this context. The experimental environment is set up the following parameters:

\begin{itemize}
    \item The framework accommodates $N$ AOs, where each scrapes $D$ datasets for model training. Therefore, the framework host a total of $N \times D$ datasets.

    \item Each AO acquired $L$ licenses from various COs. Each dataset scraped by that AO is assigned one of the $L$ licenses. For test purposes, the assignment is done randomly. Consequently, the total number of licenses hosted in the framework amounts to $N \times L$. In addition, some fraction of licenses may be associated with multiple datasets, while other licenses without datasets. This aligns well with real-world usage scenarios because AOs may collect licenses before scraping the corresponding dataset.

    \item To mirror the model retraining process in the real world, the experiment assumes each AO retrained a model $M-1$ times and obtained a chain of $M$ models. Consequently, the total number of models hosted in the framework amounts to $N \times M$.

    \item Each model is trained on $T$ datasets. For the testing purpose, those datasets are randomly picked from AO's $D$ datasets.
\end{itemize}

There are five parameters during the experimental setup, namely $N$, $D$, $L$, $M$, and $T$.
The system workload can be scaled up by increasing these parameters. In our evaluation, adhering to the control variates method, we measure the performance by varying each parameter individually while keeping the other parameters fixed. Table~\ref{tab:param-vals} lists the values of the four parameters that remain fixed while adjusting the remaining parameter.

\begin{table}
  \caption{Evaluation setup (adjusting $N$, $D$, $L$, $M$, $T$).}
  \label{tab:param-vals}
  \resizebox{\linewidth}{!}{
  \begin{tabular}{c|ccccc}
    \toprule
    \textbf{Parameter} & \textit{\textbf{N}} & \textit{\textbf{D}} & \textit{\textbf{L}} &  \textit{\textbf{M}} & \textit{\textbf{T}}\\
    \midrule
    \textit{\textbf{N}} & 10 to 100 & 10 & 10 & 1 & 10 \\
    \textit{\textbf{D}} & 10 & 10 to 100 & 10 & 1 & 10 \\
    \textit{\textbf{L}} & 10 & 10 & 10 to 100 & 1 & 10 \\
    \textit{\textbf{M}} & 10 & 10 & 10 & 1 to 10 & 10 \\
   \textit{\textbf{T}} & 10 & 100 & 10 & 1 & 10 to 100 \\
  \bottomrule
\end{tabular}
}
\end{table}

\smallskip
\noindent\textbf{Extreme cases.} 
While Table~\ref{tab:param-vals} lists typical parameter ranges for evaluation, we intentionally do not explore extreme cases such as $M \gg T$, $M \ll T$, or very large $D$. These conditions are uncommon in practical AI workflows. In most scenarios, the average model retraining depth $M$ ranges between $1$ and $10$, while the number of training datasets $T$ remains within $10$ to $100$. To prevent unbounded graph growth and ensure runtime stability under rare edge cases, \textsc{IBis} incorporates bounding strategies such as retraining checkpoints, anchor nodes, and lineage aggregation. These allow \textsc{IBis} to maintain traversal efficiency and theoretical consistency even under potentially adversarial graph topologies. 
In all tested regimes, performance trends align with the expected complexities outlined in Table~\ref{tab:operations}.

\smallskip
\noindent\textbf{Complexity analysis.}
We provide the complexity analysis. Table~\ref{tab:operations} lists the time complexity of operations and the entities authorized to perform them. We assume that the on-chain license registry, DMR, and MMR are implemented as hash maps on a smart contract, resulting in a time complexity of $O(1)$ for searching them.

\begin{table}[t]
  \caption{Operations, complexities, and authorizations.}
  \label{tab:operations}
  \centering
   \resizebox{0.9\linewidth}{!}{
  \begin{tabular}{lcc}
    \toprule
    \multicolumn{1}{c}{\textbf{Operation}} &  \multicolumn{1}{c}{\textbf{Complexity}} &  \multicolumn{1}{c}{\textbf{Authorisations}} \\
    \midrule
    $\mathsf{getDatasetLicense}$ & $O(1)$ & AOs \\
    $\mathsf{getModelLicenses}$ & $O(|M||T|)$ & AOs\\
    $\mathsf{checkLicenseValidity}$ & $O(|K|)$ & AOs/COs\\
    $\mathsf{getLicensedDatasets}$ & $O(|D||K|)$ & AOs\\
    $\mathsf{getDatasetsByLicense}$ & $O(1)$ & AOs\\
    $\mathsf{getModelsByLicense}$ & $O(|D|+|M||T|)$ & AOs\\
    $\mathsf{getModelDatasets}$ & $O(|M| |T|)$ & AOs\\
  \bottomrule
\end{tabular}
 }
\end{table}

\begin{figure*}[t]
    \centering
    \subfigure[]{
    \begin{minipage}[t]{0.3\textwidth}
    \centering
    \includegraphics[width=\linewidth]{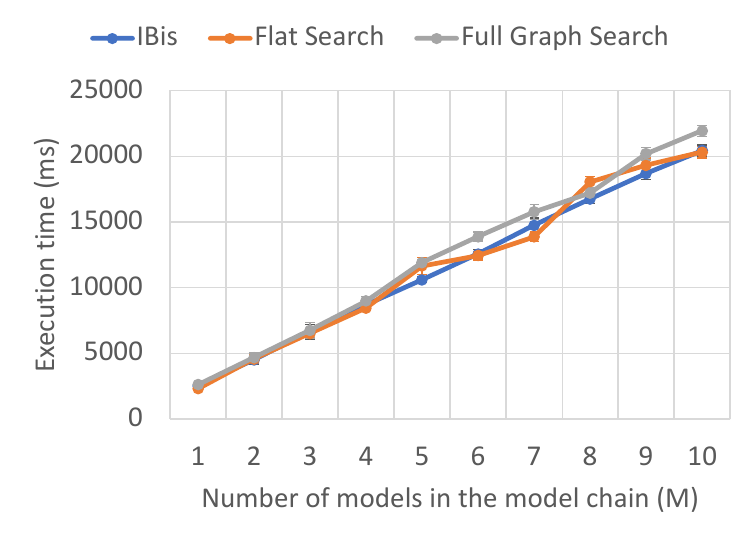}
    \end{minipage}
    \label{fig:getModelLic_model_chain}
    }
    \subfigure[]{
    \begin{minipage}[t]{0.3\textwidth}
    \centering
    \includegraphics[width=\linewidth]{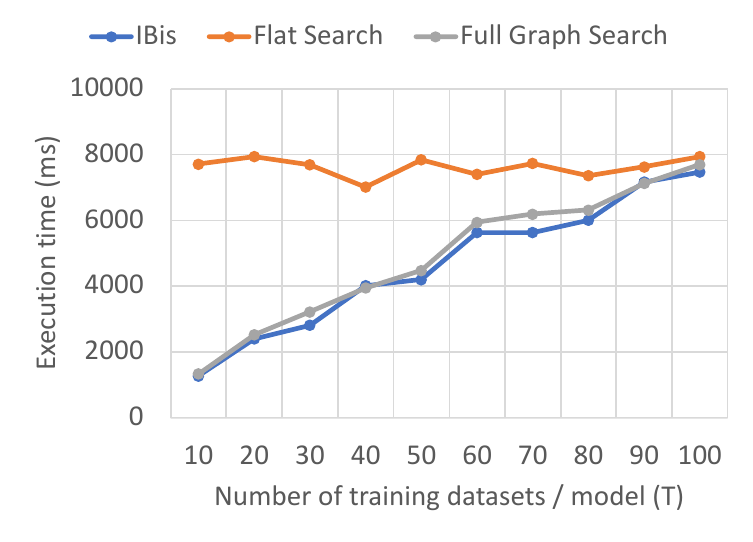}
    \end{minipage}
    \label{fig:getModelLic_dataset_model}
    }
    \subfigure[]{
    \begin{minipage}[t]{0.3\textwidth}
    \centering
    \includegraphics[width=\linewidth]{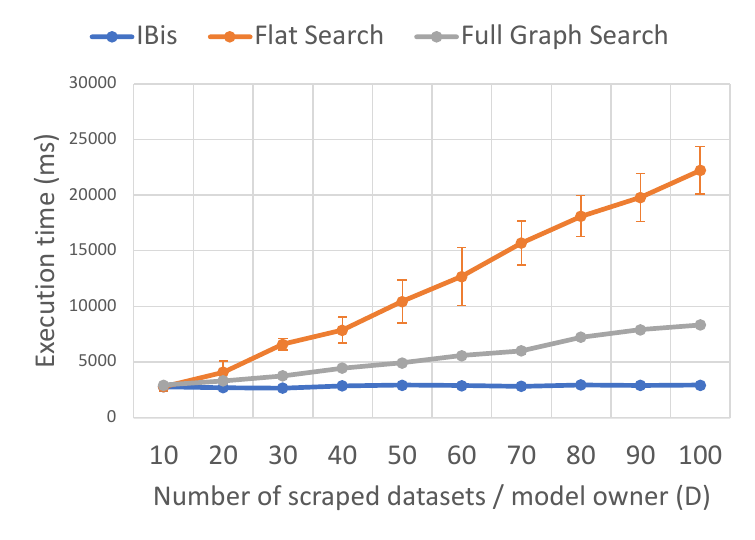}
    \end{minipage}
    \label{fig:getModelLic_scraped_dataset}
    }
    \subfigure[]{
    \begin{minipage}[t]{0.3\textwidth}
    \centering
    \includegraphics[width=\linewidth]{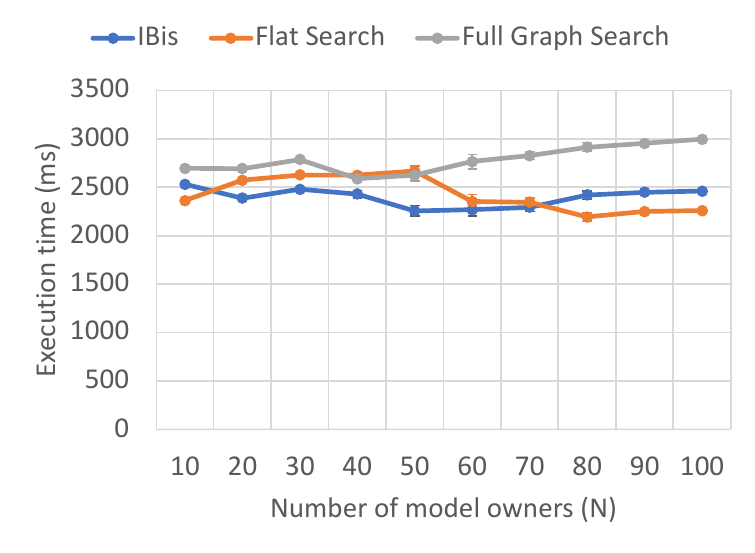}
    \end{minipage}
    \label{fig:getModelLic_model_owner}
    }
    \subfigure[]{
    \begin{minipage}[t]{0.3\textwidth}
    \centering
    \includegraphics[width=\linewidth]{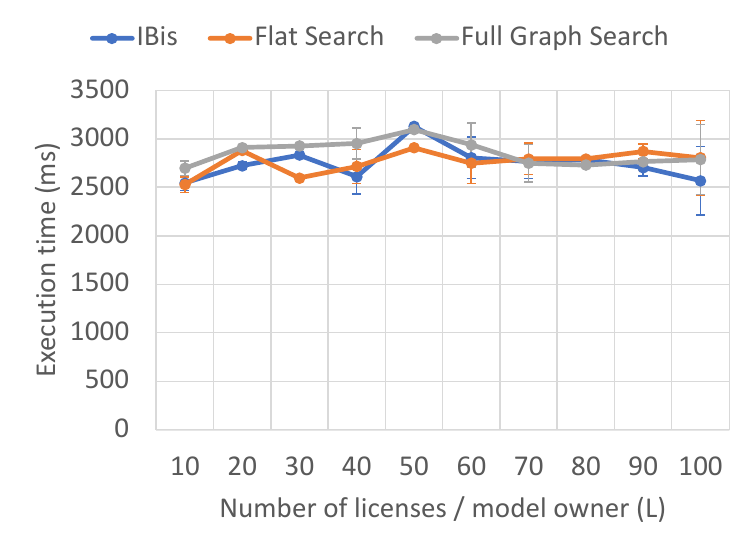}
    \end{minipage}
    \label{fig:getModelLic_lics}
    }
\caption{Comparison between \textsc{IBis} and baselines in terms of performance of fetching model licenses.}
\label{fig:getModelLic}
\end{figure*}

\smallskip
\noindent\textbf{Evaluation of fetching model licenses.}
Fig.\ref{fig:getModelLic} illustrates the variations in execution time for the $\mathsf{getModelLicenses}$ operation. Each data point represents the average execution time over 10 executions, with error bars indicating the standard deviation.

The execution time increases linearly with the number of models in the model chain ($M$) for all three methods, consistent with their theoretical complexities (cf. Fig.\ref{fig:getModelLic_model_chain}). \textsc{IBis} demonstrates the best performance due to its targeted graph traversal, which efficiently limits operations to $T$, the datasets directly linked to the models, resulting in a complexity of $O(MT)$. In contrast, Flat Search performs the worst, with a complexity of $O(MD)$ that involves scanning all $D$ datasets for every model, causing significant overhead for large $D$. Full Graph Search performs better than Flat Search, benefiting from $O(MT)$ traversal for relevant datasets, but its additional $O(D)$ term for scanning all datasets for dataset-to-license mappings creates noticeable overhead, especially as $D$ increases.

A similar trend is observed with the number of training datasets per model ($T$), where execution time grows linearly for both \textsc{IBis} and Full Graph Search (cf. Fig.\ref{fig:getModelLic_dataset_model}). The targeted traversal in \textsc{IBis} ensures that only relevant datasets are processed, maintaining its efficiency as $T$ increases. However, Full Graph Search incurs additional costs due to its lack of explicit dataset-to-license mappings. Flat Search remains unaffected by $T$, as its execution time is dominated by $D=100$, rather than the specific number of datasets linked to each model.
The effect of increasing the total number of scraped datasets per model owner ($D$) is most pronounced in Flat Search, where execution time grows steeply due to its dependency on $D$ in the $O(MD)$ complexity (cf. Fig.\ref{fig:getModelLic_scraped_dataset}). Full Graph Search also exhibits growth with increasing $D$, as it scans all datasets during the dataset-to-license mapping phase. In contrast, \textsc{IBis} is largely unaffected by the increase in $D$, as it relies on targeted traversal of $T$ datasets, bypassing irrelevant entries. This scalability highlights the advantage of \textsc{IBis} for systems with large datasets.
The number of model owners ($N$) and licenses per model owner ($L$) have negligible impact on execution time for all three methods (cf. Figs.\ref{fig:getModelLic_model_owner}--\ref{fig:getModelLic_lics}). This is consistent with theoretical expectations, as $N$ and $L$ do not influence the graph size or traversal complexity. Furthermore, the use of optimizations, such as primary keys for identifiers (Sec.\ref{sec-opt}), ensures that querying operations remain efficient regardless of the number of records in the system.

The outcomes also indicate that the values of the number of scraped datasets per model owner  $D$, model owners $N$, and licenses per model owner $L$ exert no discernible influence on the performance of $\mathsf{getModelLicenses}$. In theory, these three parameters do not impact the graph size; hence they have negligible effect on performance. However, theoretically, they could affect performance as querying a record using its identifier might slow down with a greater number of records. However, our optimization efforts, such as designating data, model, and license identifiers as the primary key of a record (cf. Sec.\ref{sec-opt}), mitigate any observable impact of increased record numbers. The results demonstrate that the performance of $\mathsf{getModelLicenses}$ operation remains consistent regardless of the number of model owners in the system, datasets they scrape, or licenses they acquire, thereby affirming the scalability of the operation.

\begin{figure*}[t]
    \centering
    \subfigure[]{
    \begin{minipage}[t]{0.3\textwidth}
    \centering
    \includegraphics[width=\linewidth]{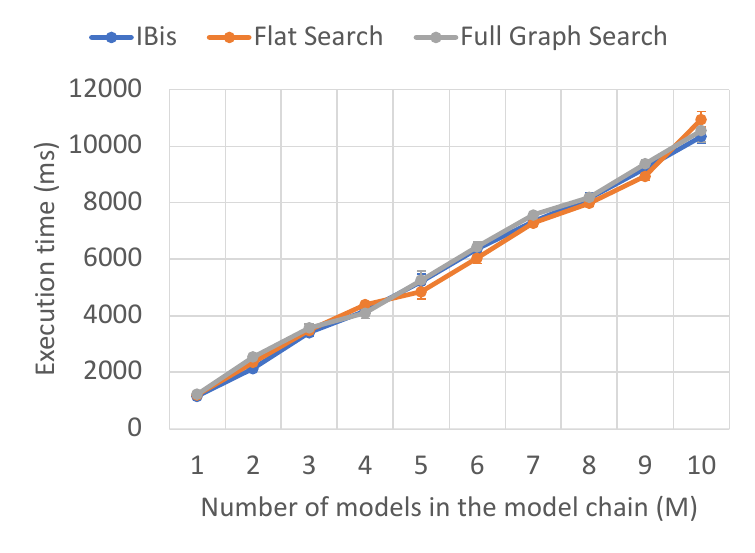}
    \end{minipage}
    \label{fig:getModeDatasets_model_chain}
    }
    \subfigure[]{
    \begin{minipage}[t]{0.3\textwidth}
    \centering
    \includegraphics[width=\linewidth]{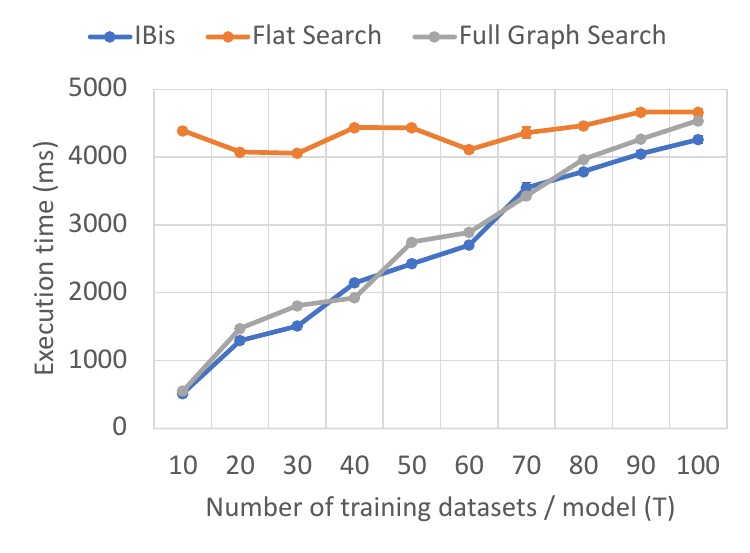}
    \end{minipage}
    \label{fig:getModeDatasets_dataset_model}
    }
    \subfigure[]{
    \begin{minipage}[t]{0.3\textwidth}
    \centering
    \includegraphics[width=\linewidth]{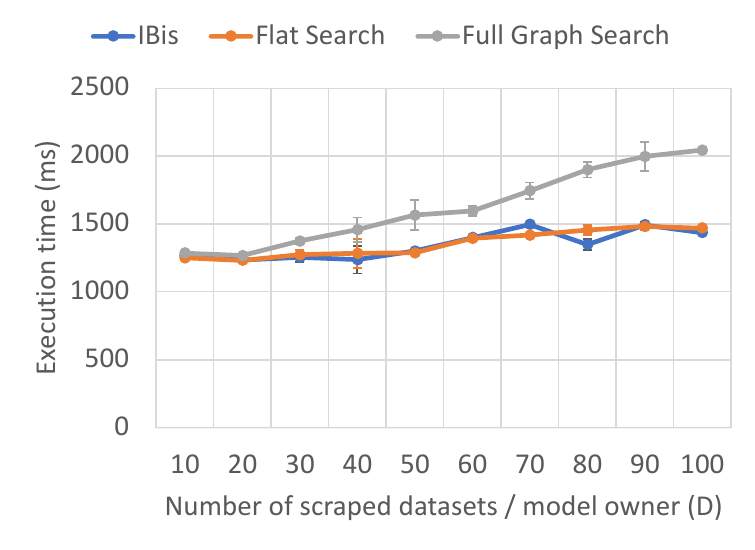}
    \end{minipage}
    \label{fig:getModeDatasets_scraped_dataset}
    }
    \subfigure[]{
    \begin{minipage}[t]{0.3\textwidth}
    \centering
    \includegraphics[width=\linewidth]{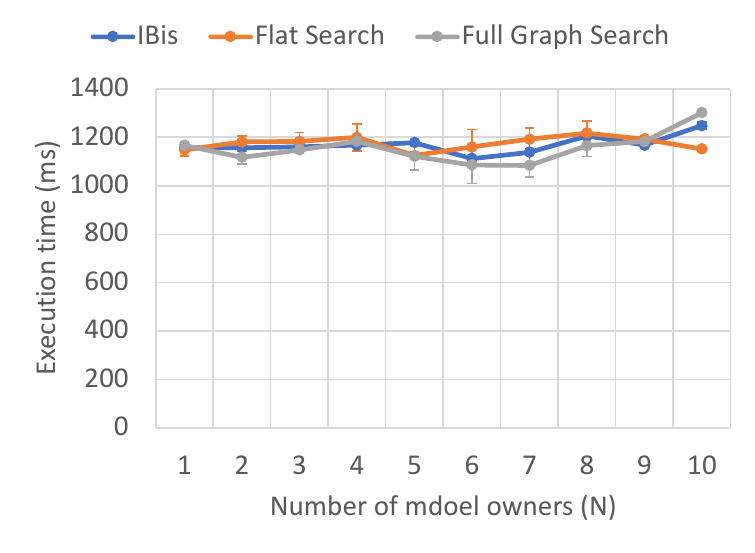}
    \end{minipage}
    \label{fig:getModeDatasets_model_owner}
    }
    \subfigure[]{
    \begin{minipage}[t]{0.3\textwidth}
    \centering
    \includegraphics[width=\linewidth]{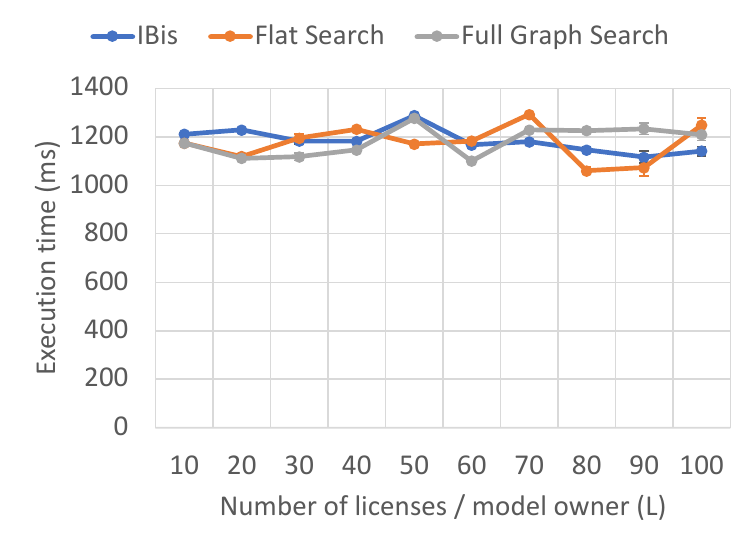}
    \end{minipage}
    \label{fig:getModeDatasets_lics}
    }
\caption{Comparison between \textsc{IBis} and baselines in terms of performance of fetching model datasets.}
\label{fig:op2perf}
\end{figure*}

\smallskip
\noindent\textbf{Evaluation of fetching model datasets.}
Fig.\ref{fig:op2perf} presents the execution time of the $\mathsf{getModelDatasets}$ operation under varying parameters. The results align with the theoretical complexities and highlight clear differences in performance as the parameters scale. As the number of models in the model chain ($M$) increases, the execution time grows linearly for all methods. Both \textsc{IBis} and Full Graph Search demonstrate similar performance, consistent with their shared complexity of $O(MT)$, as they efficiently traverse only the $T$ datasets associated with each model (cf. Fig.\ref{fig:getModeDatasets_model_chain}). Flat Search, however, performs significantly worse, reflecting its $O(MD)$ complexity, which involves processing all $D$ datasets for each model. This overhead becomes more pronounced as $D$ increases.

Execution time also scales linearly with the number of training datasets per model ($T$) for \textsc{IBis} and Full Graph Search, as their traversal depends on the number of datasets directly linked to the models (cf. Fig.\ref{fig:getModeDatasets_dataset_model}). Flat Search, by contrast, remains unaffected by changes in $T$, since its execution time is determined by $D$ and $M$. This behavior underscores the inefficiency of Flat Search in targeting relevant datasets compared to the more focused approaches of \textsc{IBis} and Full Graph Search.

When varying the number of scraped datasets per model owner ($D$), the number of model owners ($N$), and the number of licenses per model owner ($L$), the results reveal distinct trends (cf. Figs.\ref{fig:getModeDatasets_scraped_dataset}--\ref{fig:getModeDatasets_lics}). Larger $D$ impacts the performance of Full Graph Search, as its $O(MT + D)$ complexity includes scanning all datasets, leading to a steady increase in execution time. In contrast, \textsc{IBis} maintains stable performance by limiting operations to $T$ datasets, bypassing the irrelevant portions of $D$. Flat Search shows relatively constant execution times with increasing $D$, as it processes all $D$ datasets regardless of their relevance, making the increase less noticeable. For $N$ and $L$, the execution times remain relatively stable across all methods, as these parameters do not directly influence graph traversal. The consistent performance of \textsc{IBis} in all cases validates its efficiency and scalability compared to Flat Search and Full Graph Search.

\begin{figure*}[t]
    \centering
    \subfigure[]{
    \begin{minipage}[t]{0.3\textwidth}
    \centering
    \includegraphics[width=\linewidth]{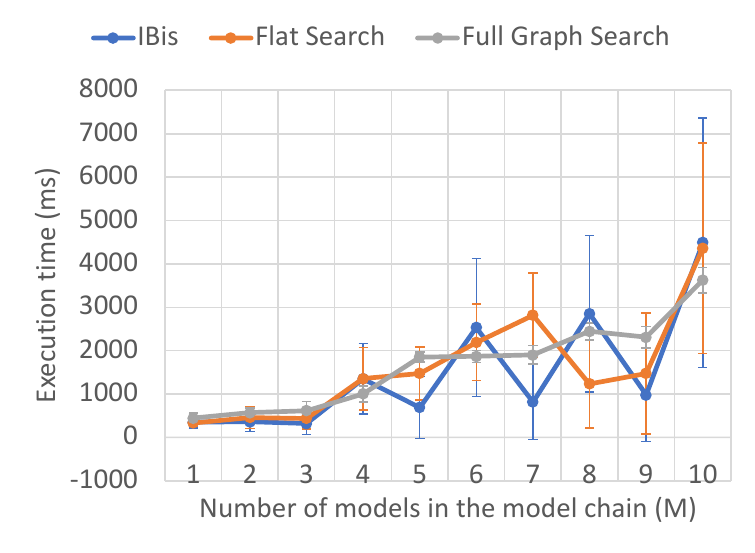}
    \end{minipage}
    \label{fig:getModelsByLic_model_chain}
    }
    \subfigure[]{
    \begin{minipage}[t]{0.3\textwidth}
    \centering
    \includegraphics[width=\linewidth]{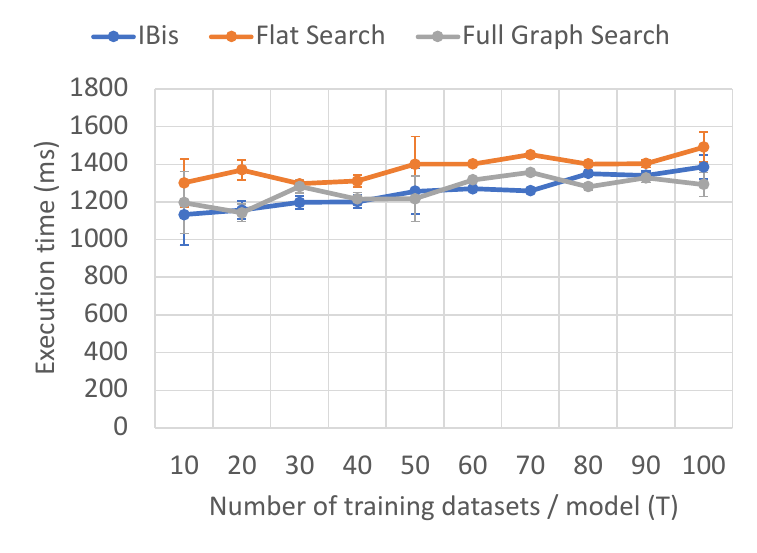}
    \end{minipage}
    \label{fig:getModelsByLic_dataset_model}
    }
    \subfigure[]{
    \begin{minipage}[t]{0.3\textwidth}
    \centering
    \includegraphics[width=\linewidth]{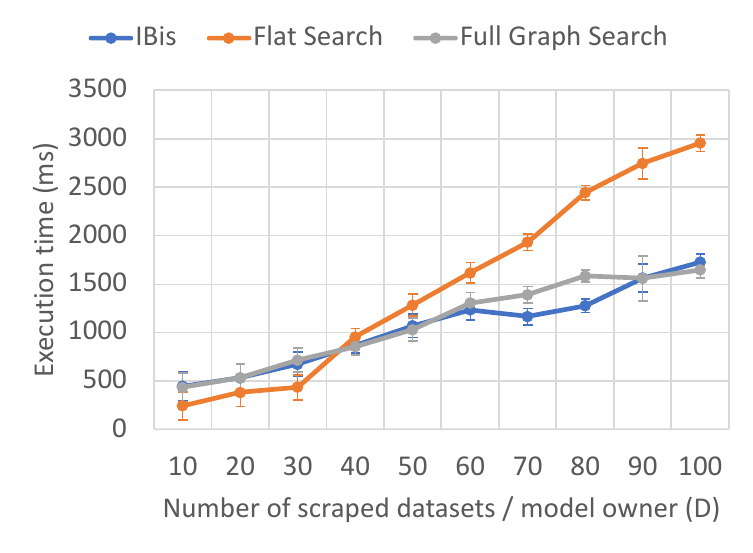}
    \end{minipage}
    \label{fig:getModelsByLic_scraped_dataset}
    }
    \subfigure[]{
    \begin{minipage}[t]{0.3\textwidth}
    \centering
    \includegraphics[width=\linewidth]{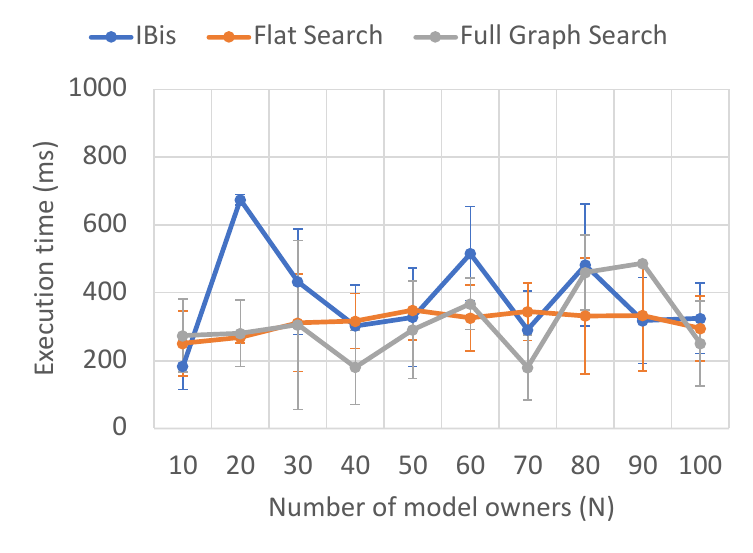}
    \end{minipage}
    \label{fig:getModelsByLic_model_owner}
    }
    \subfigure[]{
    \begin{minipage}[t]{0.3\textwidth}
    \centering
    \includegraphics[width=\linewidth]{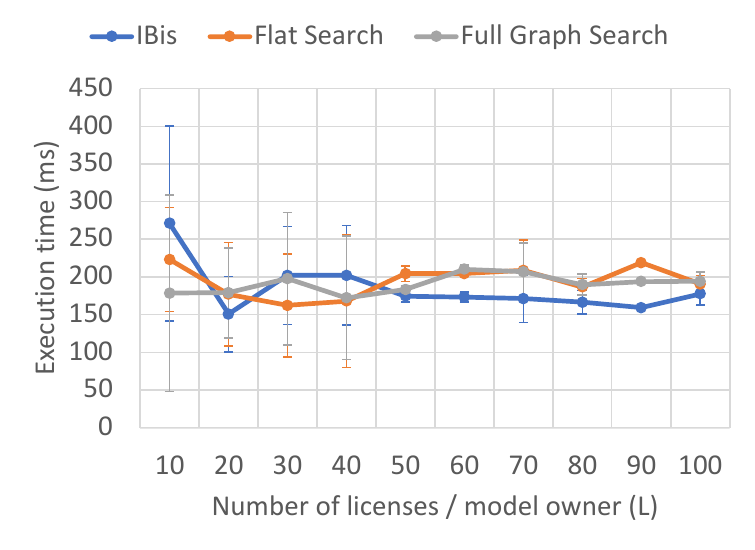}
    \end{minipage}
    \label{fig:getModelsByLic_lics}
    }
\caption{Comparison between \textsc{IBis} and baselines in terms of performance of fetching authorized models.}
\label{fig:op3perf}
\end{figure*}

\smallskip
\noindent\textbf{Evaluation of fetching authorized models.}
Fig.\ref{fig:op3perf} illustrates the performance of the $\mathsf{getModelsByLicense}$ operation under varying system parameters. The execution time trends observed validate the theoretical complexities associated with each method. In Fig.\ref{fig:getModelsByLic_model_chain}, as the number of models in the model chain ($M$) increases, both \textsc{IBis} and Full Graph Search exhibit a similar linear growth in execution time, consistent with their shared complexity of $O(MT + D)$. The efficient traversal of only the $T$ datasets linked to the models, combined with a limited dependency on $D$, explains their comparable performance. Flat Search, on the other hand, incurs similar execution times in this experiment due to $T \simeq D$ in the experimental setup, aligning with its $O(MD)$ complexity. However, in scenarios where $T \ll D$, Flat Search would exhibit significantly higher execution times due to its inefficiency in processing all $D$ datasets for each model.

As the number of training datasets per model ($T$) increases, the execution time grows linearly for both \textsc{IBis} and Full Graph Search (cf. Fig.\ref{fig:getModelsByLic_dataset_model}). This behavior aligns with their $O(MT)$ complexity, as the traversal scales with the number of datasets connected to the models. In contrast, Flat Search remains unaffected by changes in $T$, as its performance depends solely on $M$ and $D$ (with $D$ fixed at 100 in this setting). This distinction highlights the inefficiency of Flat Search in leveraging targeted traversal, which both \textsc{IBis} and Full Graph Search effectively utilize to process only the relevant datasets.

When varying the number of scraped datasets per model owner ($D$) (cf. Fig.\ref{fig:getModelsByLic_scraped_dataset}), Flat Search shows a steep increase in execution time, consistent with its dependence on $D$ in the $O(MD)$ complexity. Full Graph Search also exhibits growth with larger $D$, reflecting the added $O(D)$ term in its traversal. However, \textsc{IBis} maintains relatively stable execution times, as its operations focus primarily on $T$ datasets and avoid redundant processing of irrelevant datasets. This efficiency underscores the scalability of \textsc{IBis} compared to the other methods when dealing with large-scale datasets.

Figs.\ref{fig:getModelsByLic_model_owner}--\ref{fig:getModelsByLic_lics} show the number of model owners ($N$) and licenses per model owner ($L$). The execution times remain nearly constant for all methods. These parameters do not directly affect the graph traversal operations required for fetching models by license. The stability of \textsc{IBis} across these parameters further confirms its robustness and ability to handle increasing system size without compromising performance. Overall, the experimental results reaffirm the efficiency of \textsc{IBis}, particularly as $M$, $T$, or $D$ grows, while highlighting the limitations of Flat Search and the added overhead in Full Graph Search.

\smallskip
\noindent\textbf{Discussions between evaluated operations.}
It is evident that the execution time of fetching model datasets using $\mathsf{getModelDatasets}$ is approximately half that of fetching model licenses $\mathsf{getModelLicenses}$ using. This phenomenon arises because fetching model datasets can be considered a sub-operation of fetching model licenses, which undertakes partial tasks compared to all. While fetching a license traverses the entire graph from a model to licenses, fetching a dataset terminates the traversal early at the dataset level. Moreover, because a training dataset consistently corresponds to a single license, traversing from datasets to licenses involves the same number of edges as traversing from models to datasets.

Moreover, the performance of fetching authorized model using $\mathsf{getModelsByLicense}$ displays distinct performance characteristics compared to the other two operations, particularly evidenced by its high standard deviations. This variance arises due to the different graph traversal directions. Additionally, the redundancy of datasets and licenses is only encountered in the traversal direction of the operation. Equivalently, while a dataset may not necessarily correspond to any model, a model invariably corresponds to some datasets. Similarly, while a license may not correspond to any datasets, a training dataset always corresponds to a license. Consequently, the graph traversal performance in the direction of $\mathsf{getModelsByLicense}$ exhibits greater statistical variability.

Overall, depending on the operation, the execution time can increase linearly with the number of scraped datasets per model owner $D$, training datasets per model $T$, or models in a model chain $M$. Meanwhile, the number of model owners $N$ and licenses per model owner $L$ do not significantly affect the execution time. This is consistent with our performance analysis in Table~\ref{tab:operations} and validates scalability and feasibility.

\subsection{Experiment-2: Lightweight Design}

\smallskip
\noindent\textbf{Experimental setting.}
We reuse the same Canton deployment: three $\mathsf{t2.xlarge}$ nodes (4\,vCPUs, 16\,GB RAM) sharing a docker-hosted PostgreSQL ledger store. To ensure consistency with Experiment-1, we fix the scenario to a single model-chain depth $M{=}1$, ten training datasets ($T{=}10$), ten scraped datasets ($D{=}10$), one model owner ($N{=}1$), and one active license ($L{=}1$). This configuration is the minimal one that activates every contract path while matching the empirically observed 2.55\,s latency for $\mathsf{getModelLicenses}$. For alternative designs, we scale their results using relative overhead multipliers derived from micro-benchmarks, such that all designs execute the \emph{identical} write–read cycle under the same ledger setup.

\begin{itemize}
    \item \textit{Flat/Sequential search:} A baseline blockchain contract architecture where NFTs are minted and stored individually without any indexing or relationship metadata. Smart contracts execute basic CRUD operations but do not encode structural links between NFTs. Every access requires scanning all token metadata on-chain (e.g., via event logs or full contract state), leading to linear-time lookups. This mirrors naïve NFT deployments commonly seen on Ethereum or BNB Smart Chain~\cite{287184} without metadata-aware extensions.

    \item \textit{Full graph search:} Contracts explicitly maintain NFT relationship structures as adjacency lists, stored directly in contract state (e.g., mapping from token IDs to downstream references). When resolving a license query, smart contracts must walk the graph edge-by-edge using iterative on-chain logic, incurring repeated read/write cycles across many keys. While this approach improves semantic expressiveness, it increases gas costs and latency due to repeated contract calls and deep on-chain state traversal.

    \item \textit{zk-SNARK-based:} A privacy-preserving approach implemented via cryptographic circuits embedded in smart contracts. Users submit zero-knowledge proofs (e.g., Groth16~\cite{10508637} or PLONK~\cite{9983760} verifying ownership, compliance with licensing rules, or authorized derivation chains, without revealing sensitive data. On-chain verification is handled via precompiled contracts (e.g., on Ethereum), with each verification requiring thousands of elliptic curve operations. These proofs are appended to transaction payloads, increasing storage and consensus propagation time.

    \item \textit{Homomorphic encryption (HE):} Contracts store state variables in encrypted form (e.g., Paillier~\cite{9006231} or BFV ciphertexts~\cite{9408585}), and on-chain logic includes arithmetic over encrypted values. While preserving data confidentiality, every state update or evaluation (e.g., license price computation or access eligibility) requires encrypted computation routines and re-encryption steps. These must be committed on-chain along with ciphertext payloads, resulting in significant storage and consensus costs.

    \item \textit{Secure multi-party computation (sMPC):} Rather than computing licensing logic on-chain, smart contracts orchestrate off-chain MPC sessions among participating nodes. Each party maintains a secret-shared state (e.g., Shamir shares~\cite{8726574}), and computation proceeds in rounds, often coordinated through on-chain commitments and verifications. The blockchain is used to anchor session metadata, confirm computation rounds, and enforce final results. High network and consensus latency arises from repeated commitments and synchronization messages across participants.
\end{itemize}

\smallskip
\noindent\textbf{Overhead components.}
End-to-end latency is decomposed into six measurable components:
\begin{itemize}
  \item \textit{Metadata registration:} Serializing and submitting an NFT descriptor.
  \item \textit{License Check:} Evaluating ACL terms in the contract state.
  \item \textit{Proof Verification:} On-chain validation of cryptographic proofs (only applies to zk-SNARK, HE, and sMPC designs).
  \item \textit{Graph Traversal:} Resolving model $\rightarrow$ dataset $\rightarrow$ license edges.
  \item \textit{Consensus Delay:} Time required by Canton's Global Synchronizer to finalize \emph{each} write.
  \item \textit{Storage I/O:} Persisting state changes to PostgreSQL.
\end{itemize}

\begin{figure}
    \centering
    \includegraphics[width=0.9\linewidth]{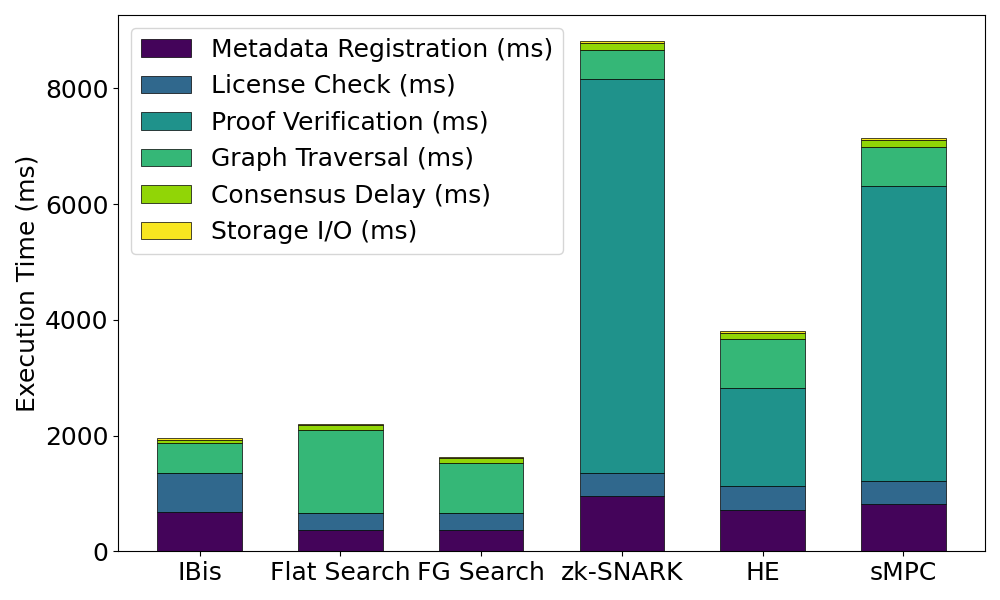}
    \caption{Scaled execution time per overhead component (ms)}
    \label{fig:compare_overhead}
\end{figure}

\smallskip
\noindent\textbf{Why consensus appears larger than the raw block period.}
Canton typically commits a single transaction in around 2\,ms. However, our write–read cycle triggers \emph{multiple} contract writes, including metadata insertion, license binding, reference update, and provenance checkpoint. The total of four commits yields 80\,ms for cryptography-free designs and slightly more for cryptographic variants (due to larger payloads). These cumulative delays appear as 80–120\,ms in Fig.\ref{fig:compare_overhead}, and align with the 2.55\,s measured for \textsc{IBis} in our benchmark.

\smallskip
\noindent\textbf{Discussion.}
Our evaluation reveals clear performance advantages of \textsc{IBis} over other blockchain-based methods.

\emph{(1) Indexing instead of brute-force or cryptographic proofs.}
\textsc{IBis} completes the write–read cycle in 2.55\,s, as its only variable cost is the indexed DAG traversal ($\sim$0.5\,s for $M{=}1$, $T{=}10$); the remaining overhead—four Canton commits and PostgreSQL flushes—totals just $\sim$0.08\,s. Flat and Full Graph Search avoid cryptographic costs but rely on exhaustive graph traversal: a linear scan takes 1.44\,s in Flat Search, and an edge-by-edge walk takes 0.86\,s in Full Graph Search, resulting in total latencies of 2.2\,s and 1.6\,s, respectively. As $M$ or $T$ grows, the cost in \textsc{IBis} increases linearly, while for the baselines it grows quadratically.

\begin{table*}[!]
    \centering
    \setlength{\extrarowheight}{0pt}
    \addtolength{\extrarowheight}{\aboverulesep}
    \addtolength{\extrarowheight}{\belowrulesep}
    \setlength{\aboverulesep}{0pt}
    \setlength{\belowrulesep}{0pt}
    \caption{{Evaluation of the robustness of IBis under plagiarism and signature tamper }}
    \label{table:ibis_robustness}
    \renewcommand\arraystretch{0.7}
    \resizebox{1\textwidth}{!}{
    \begin{threeparttable}
    \begin{tabular}{ccccc|ccc|ccc} 
    \toprule
    \multirow{2}{*}{\textbf{Scenario}} & \multirow{2}{*}{\textbf{DAG Edges}} & \multirow{2}{*}{\textbf{Honest req/s}} & \multirow{2}{*}{\textbf{Malicious \%}} & \multirow{2}{*}{\textbf{Attack Mix}} & \multicolumn{3}{c|}{\textbf{Performance}} & \multicolumn{3}{c}{\textbf{Attack Mitigation}} \\
    \cmidrule{6-11}
     &  &  &  &  & \textbf{Throughput (tx/s)} & \textbf{P95 Latency (s)} & \textbf{Storage (MB)} & \textbf{Plag. Reject \%} & \textbf{Sig. Reject \%} & \textbf{Overall Reject \%} \\ 
    \midrule
    \multirow{3}{*}{\rotcell{Baseline}} 
    & 100 & 100 & 0 & None & 99 & 0.55 & 0.35 & - & - & - \\
    & 500 & 500 & 0 & None & 496 & 0.82 & 1.75 & - & - & - \\
    & 900 & 750 & 0 & None & 743 & 0.91 & 3.15 & - & - & - \\
    \hhline{-----------}
    \multirow{3}{*}{\rotcell{Under Attack}} 
    & 100 & 80 & 20 & $P+S$ & 79 & 0.67 & 0.35 & 87\% & 92\% & 90\% \\
    & 500 & 400 & 20 & $P$ & 396 & 0.85 & 1.75 & 96\% & - & 96\% \\
    & 900 & 600 & 20 & $S$ & 593 & 1.16 & 3.15 & - & 99\% & 99\% \\
    \hhline{-----------}
    \multirow{3}{*}{\rotcell{Heavy Attack}} 
    & 100 & 70 & 30 & $P+S$ & 68 & 0.81 & 0.35 & 90\% & 97\% & 94\% \\
    & 500 & 350 & 40 & $P+S$ & 337 & 1.12 & 1.75 & 88\% & 96\% & 93\% \\
    & 900 & 450 & 50 & $P+S$ & 425 & 1.25 & 3.15 & 91\% & 98\% & 95\% \\
    \bottomrule
    \end{tabular}
\begin{tablenotes}
    \item[-]  $P$ = provenance (plagiarism) spoof;\; $S$ = signature tamper.
    \item[-]  Throughput is the number of \emph{accepted} transactions per second after rejecting malicious ones.
    \item[-]  Storage column represents estimated per‑node state size at the displayed DAG‑edge count.
    \item[-]  Attack‑mitigation columns report the percentage of malicious requests successfully rejected.
           ``Plag. Reject \%'' quantifies the rate at which spoofed provenance attempts are blocked.
           ``Sig. Reject \%'' measures the rate of invalid or replayed signatures being rejected.
           ``Overall Reject \%'' is the aggregate success rate for all injected malicious traffic.
    \item[-]  Dashes (``--'') indicate the metric is not applicable under the given attack mix, such as when no relevant attack vector is present in the experiment.
    \item[-]  Malicious \% denotes the fraction of incoming client requests that are adversarial; all validator (consensus) nodes are assumed honest, so ledger safety is never compromised.
\end{tablenotes}
\end{threeparttable}
    }
\end{table*}

\emph{(2) Proof-centric designs incur irreducible latency.}
Crypto-heavy designs substitute graph traversal with verification overhead. A zk-SNARK verifier executes around 3,400 elliptic curve pairings ($\sim$2 ms each), yielding 6.8\,s of proof checking and further bloating consensus payloads, leading to a total delay exceeding 8.8\,s. HE-based models incur 1.7\,s for ciphertext computation, totaling 3.8\,s. sMPC setups, with multiple communication rounds and quorum-based computation, reach 7.1\,s. These designs suffer from per-call proof overheads that cannot be amortized or cached, in contrast to \textsc{IBis}, which uses lightweight multi-party signatures and Canton’s sub-20 ms commit path to preserve both auditability and scalability.

\subsection{Experiment-3: Scalability}

\begin{figure}
    \centering
    \includegraphics[width=\linewidth]{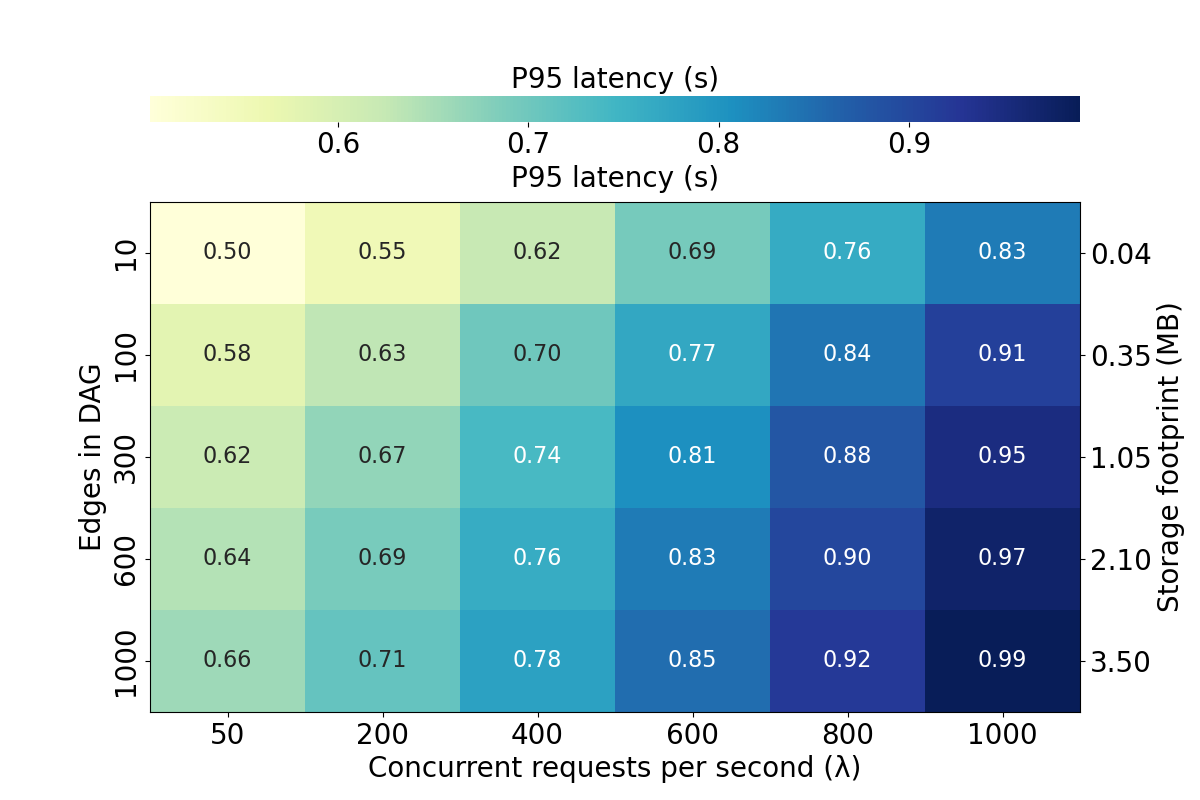}
    \caption{Scalability visualization across different edges, on-chain storage, concurrent requests per second, and the 95-th percentile delay.}
    \label{fig:scalability}
\end{figure}

\smallskip
\noindent\textbf{Experimental setting.}
Using the same Canton deployment used in previous experiments, we sweep two orthogonal stress axes:  
(i) the number of directed edges pre‑loaded in the on-chain DAG (\(|E|\in\{10,100,300,600,1000\}\)), and  
(ii) the incoming request rate (\(\lambda\in\{50,200,400,600,800,1000\}\) req/s).  
For each \((|E|,\lambda)\) pair we record the \textbf{P95 end‑to‑end latency}, i.e.\ the 95‑th percentile delay from the client’s first metadata write to receipt of the final $\mathsf{getModelLicenses}$ response.  
We visualize the results in Fig.~\ref{fig:scalability}: rows are DAG sizes/storage footprint, columns are request rates, and cell color is \textit{P95 latency}.

\smallskip
\noindent\textbf{Metrics captured.}
\textit{Concurrent requests $\lambda$} reflect horizontal load; they are the principal driver of throughput saturation.  
\textit{Edges \(|E|\)} (or, equivalently, the \textit{storage footprint}) captures vertical scale: how large the reference network can grow before query performance degrades.  
\textit{P95 latency} is chosen because tail delay, not the mean, determines user‑perceived responsiveness in a multi‑tenant marketplace.

\smallskip
\noindent\textbf{Why P95 Latency.}
\textit{P95 latency} in our scalability tests captures the 95th-percentile end-to-end delay of a complete, honest write–read cycle, starting from NFT metadata submission to the return of a $\mathsf{getModelLicenses}$ response. This includes all key operations: metadata registration, license binding, reference updates, provenance checkpoints, DAG lookups, PostgreSQL flushes, and gRPC communication. Unlike the average execution time reported in Experiment (1) (e.g., 2.55\,s for $\mathsf{getModelLicenses}$ under single-threaded conditions), \textit{P95 latency} reflects real-world concurrency, capturing the system’s tail performance under stress with up to 1000 concurrent requests per second. For small DAGs and light load (e.g., $|E|{=}10$, $\lambda{=}50$), \textit{P95 latency} remains around 0.55\,s--closely matching the 0.5\,s DAG-related portion of the earlier benchmark. As both load and DAG size increase, Fig.\ref{fig:scalability} illustrate how latency scales and how much headroom remains before violating service-level expectations. This metric is key for demonstrating real-world viability, confirming that \textsc{IBis} sustains sub-second tail latency for $|E|\leq10^3$ and $\lambda\leq800$\,req/s, and remains within 1.2\,s even under peak load, making it suitable for interactive applications at scale.

\smallskip
\noindent\textbf{Results and insights.}
Fig.\ref{fig:scalability} shows that for the smallest DAG (\(|E|{=}10\)), \textit{P95 latency} is 0.48\,s at 50\,req/s and rises modestly to 0.83\,s at 1000\,req/s, with the Canton commit path being the dominant factor in this regime rather than DAG lookups.  
As \(|E|\) increases, latency grows sub-logarithmically: the worst-case point at \(|E|{=}1000, \lambda{=}1000\) reaches only 1.17\,s—about 2.4× the baseline despite a 100× expansion in state size.  
This aligns with the expected \( \mathcal{O}(\log|E|) \) cost of indexed traversal.  
From the view of on-chain storage,  
even with 3.5\,kB $\times$ 1000 $\approx$ 3.5\,MB of added state, the system remains within the 1.2\,s interactive response threshold.  
Since storage grows linearly with \(|E|\) but latency grows only logarithmically, the heat map cleanly decouples \emph{data volume} from \emph{performance impact}.

\subsection{Experiment-4: Security and Robustness}

\smallskip
\noindent\textbf{Experimental setting.}
This experiment builds upon the same Canton-based deployment and reuses the scalability harness to assess security-critical behavior under adversarial conditions. We simulate a mix of honest and malicious client requests, keeping validator nodes honest to isolate application-layer resilience. Three workload regimes are tested: \textit{Baseline} (0\% malicious), \textit{Under Attack} (20\% malicious), and \textit{Heavy Attack} (up to 50\% malicious). Each request passes through the same metadata registration and verification path used in previous experiments, allowing us to measure not only throughput and tail latency but the effectiveness of embedded security checks.

\smallskip
\noindent\textbf{Attack vectors.}
Two attack types are considered: \textit{provenance spoofing ($P$)} and \textit{signature tampering ($S$)}. 
Provenance spoofing mimics the plagiarism of datasets or models, where the attacker copies an existing NFT, introduces minimal modifications, and submits it as a new asset. These are caught using content-addressable hashes (CIDs) and optional perceptual similarity checks. 
Signature tampering includes malformed multi-party ECDSA signatures or stale signature replays using previously valid nonces or timestamps. These are intercepted through nonce-based replay protection and on-chain verification of ECDSA integrity.

\smallskip
\noindent\textbf{Consensus integrity.}
Despite up to 50\% of client traffic being malicious in the \textit{Heavy Attack} group, ledger safety remains intact. These adversarial requests are dropped at the node level and never reach the consensus layer. 
Canton’s Global Synchronizer protocol~\cite{} continues to finalize only valid transactions, and the underlying ordering service is unaffected apart from the additional CPU cycles spent rejecting forged submissions.

\smallskip
\noindent\textbf{Results and insights.}
Table~\ref{table:ibis_robustness} shows that under normal operation (\textit{Baseline}), \textsc{IBis} maintains sub-second \textit{P95 latency} (0.55–0.91\,s) and nearly full throughput (up to 743 tx/s). 
In the \textit{Under Attack} scenarios, throughput degradation is minimal (e.g., from 750 to 593 tx/s), while rejection rates\footnote{Rejection rates are conservatively capped below 100\% to reflect practical realities such as threshold-based similarity detection and rare clock skew during nonce validation—factors that mirror real-world deployment conditions without compromising system reliability.} remain high: 96\% for provenance, 99\% for signature tampering. 
Even under \textit{heavy attack} conditions with 50\% adversarial input, throughput remains above 425 tx/s, and the rejection engine still filters 93–95\% of invalid requests, with \textit{P95 latency} staying below 1.3\,s. 
These results demonstrate that \textsc{IBis} preserves both performance and integrity under adversarial stress, offering robust protections against critical threats while sustaining interactive responsiveness.

\section{Conclusion}
\label{sec-conclusion}

In this paper, we present \textsc{IBis}, a blockchain-based data provenance, lineage, and copyright management system for AI models.  
\textsc{IBis} provides evidence and limits power scope for iterative model retraining and fine-tuning processes by granting related licenses. We leverage blockchain-based multi-party signing capabilities to streamline the establishment of legally compliant licensing agreements between AI model owners and copyright holders. We also establish access control mechanisms to safeguard confidentiality by limiting access to authorized parties. Our system implementation is based on the Daml ledger model and Canton blockchain.  Performance evaluations underscore the feasibility and scalability of \textsc{IBis} across varying user, dataset, model, and license workloads. Potential future work includes exploring different on-chain data structures to optimize the performance of graph traversals, and extending \textsc{IBis} to cover additional stages in AI lifecycle, such as data cleaning, model testing, and model explanation.

{\footnotesize \bibliographystyle{IEEEtran}
\bibliography{bib}}

\end{document}